\documentclass{aa}

\usepackage[varg]{txfonts}
\usepackage[utf8]{inputenc}

\usepackage{hyperref}
\hypersetup{
	colorlinks = true,
	citecolor=blue,
	urlcolor=purple,
	linkcolor=magenta
}
\usepackage[dvipsnames]{xcolor}
\usepackage{graphicx}
\usepackage{amsmath}
\usepackage{amssymb}
\usepackage{afterpage} 

\usepackage{booktabs}
\usepackage[flushleft]{threeparttable}

\usepackage{natbib,twoopt}
\bibpunct{(}{)}{;}{a}{}{,} 
\usepackage{CJK}
\graphicspath{{./}{figures/}}
\begin{document}
\begin{CJK*}{UTF8}{gbsn}

\title{Stellar Chromospheric Activity Database of Solar-like Stars Based on the LAMOST Low-Resolution Spectroscopic Survey: II. the bolometric and photospheric calibration}


\author{Weitao Zhang\inst{1}
        \and
        Jun Zhang\inst{1}
        \and
        Han He\inst{2,3}
        \and
        Ali Luo\inst{2,3,4}
        \and
        Haotong Zhang\inst{2,3,4}
        }
  
\institute{School of Physics and Optoelectronics Engineering, Anhui University, Hefei 230601, China; \email{zjun@ahu.edu.cn}
\and
National Astronomical Observatories, Chinese Academy of Sciences, Beijing 100101, China; \email{hehan@nao.cas.cn}
\and
University of Chinese Academy of Sciences, Beijing 100049, China
\and
CAS Key Laboratory of Optical Astronomy,  Chinese Academy of Sciences, Beijing 100101, China}

  \titlerunning{Stellar Chromospheric Activity Database of Solar-like Stars: II.}
  \authorrunning{Zhang et al.}
  
\abstract
{The dependence of stellar magnetic activity on stellar parameters would be inspired by the chromospheric activity studies based on the large-scale spectroscopic surveys.}
{The main objective of this project is to provide the chromospheric activity parameters database for the LAMOST Low-Resolution Spectroscopic Survey (LRS) spectra of solar-like stars and explore the overall property of stellar chromospheric activity.}
{The \ion{Ca}{II} H and K lines are employed to construct indicators for assessing and studying the chromospheric activity of solar-like stars.	
We investigate the widely used bolometric and photospheric calibrated chromospheric activity index $R'_{\rm HK}$, derived from the method in the classic literature ($R'_{\rm HK,classic}$) and the method based on the PHOENIX model ($R'_{\rm HK,PHOENIX}$).
Since the detailed stellar atmospheric parameters, effective temperature ($T_{\rm eff}$), surface gravity ($\log\,g$), and metallicity ([Fe/H]), are available for LAMOST, we estimate the chromospheric activity index $R'_{\rm HK,PHOENIX}$, along with the corresponding bolometric calibrated index $R_{\rm HK,PHOENIX}$, taking these parameters into account.}
{We provide the database of the derived chromospheric activity parameters for 1,122,495 LAMOST LRS spectra of solar-like stars.
Our calculations show that $\log\,R'_{\rm HK,PHOENIX}$ is approximately linearly correlated with $\log\,R'_{\rm HK,classic}$.
The results based on our extensive archive support the view that the dynamo mechanism of solar-like stars is generally consistent with the Sun; and the value of solar chromospheric activity index is located at the midpoint of the solar-like star sample.
We further investigate the proportions of solar-like stars with different chromospheric activity levels (very active, active, inactive and very inactive).
The investigation indicates that the occurrence rate of high levels of chromospheric activity is lower among the stars with effective temperatures between $5600$ and $5900 \,{\rm K}$.
}
{}


\maketitle

\section{Introduction}\label{sec:introduction}

Stellar chromospheric activity, known as the performance of stellar magnetic activity, is expected to reveal the physical mechanism of stars \citep{2008LRSP....5....2H}.
The emission in the line cores of \ion{Ca}{II} H and K lines is commonly recognized to be sensitive to stellar chromospheric activity.
An empirical chromospheric activity index $S_{\rm MWO}$ was introduced to quantify the emission of \ion{Ca}{II} H and K lines observed in the Mount Wilson Observatory (MWO) \citep{1968ApJ...153..221W, 1978PASP...90..267V}.
Since $S_{\rm MWO}$ is defined as the ratio between the emission flux in the line cores of \ion{Ca}{II} H and K lines and the pseudo-continuum flux (the flux of two 20\,{\AA} reference bands in the violet and red sides), it is concise and effective for characterizing the stellar activity cycle \citep{1978ApJ...226..379W}.
However, $S_{\rm MWO}$ is related to the continuum flux which is governed by the stellar effective temperature (or equivalently, the color index) \citep{1982A&A...107...31M}.
As a result, it would be inflexible for comparing the emission of \ion{Ca}{II} H and K lines among stars of different spectral types.

The ratio between the stellar surface flux in the line core of \ion{Ca}{II} H and K lines and the stellar bolometric flux, denoted as $R_{\rm HK}$, is considered to be marginally affected by the stellar effective temperature (or the color index) and can be derived from $S_{\rm MWO}$ \citep{1979ApJS...41...47L, 1982A&A...107...31M, 1984A&A...130..353R}.
\citet{1982A&A...107...31M} and \citet{1984A&A...130..353R} introduced the bolometric factor $C_{\rm cf}$ (depends on the color index $B-V$) and the factor $K$ to convert $S_{\rm MWO}$ to the stellar surface flux in the line cores of \ion{Ca}{II} H and K lines.
Meanwhile, the photospheric fluxes contained in the line cores of \ion{Ca}{II} H and K lines could not be ignored, especially for solar-like stars \citep{1984ApJ...276..254H, 1984ApJ...279..763N}.
The photospheric contribution $R_{\rm phot}$, which represents the photospheric flux normalized by the stellar bolometric flux, can analogously be deduced as a function of $B-V$ \citep{1984ApJ...279..763N}.
Subtracting $R_{\rm phot}$ from $R_{\rm HK}$, one can derive the widely used bolometric and photospheric calibrated chromospheric activity index $R'_{\rm HK}$. 

The $R'_{\rm HK}$ is frequently employed to characterise the relationships between stellar chromospheric activity and other stellar properties such as rotation period \citep{1984ApJ...279..763N, 2015MNRAS.452.2745S, 2017A&A...600A..13A, 2022MNRAS.514.4300B, 2022ApJ...929...80B} and stellar age \citep{2008ApJ...687.1264M, 2013A&A...551L...8P, 2018A&A...619A..73L, 2020MNRAS.491..455B}.
The derivation of $R'_{\rm HK}$ may be influenced by the bolometric factor $C_{\rm cf}$, the value of $K$, the photospheric contribution $R_{\rm phot}$ and $S_{\rm MWO}$.
\citet{2007A&A...469..309C} compared the relationship between the ${\rm H}\alpha$ line and the \ion{Ca}{II} H and K lines, where they recalibrated the $C_{\rm cf}$ to the range of $0.45 \le B-V \le 1.81$.
\citet{2015MNRAS.452.2745S} and \citet{2017A&A...600A..13A} concentrated on the relationship of $R'_{\rm HK}$ and the rotation period for M dwarfs.
\citet{2015MNRAS.452.2745S} extended the bolometric factor $C_{\rm cf}$ and the photospheric contribution $R_{\rm phot}$ to $B-V$  = 1.90 using the empirical spectral library. \citet{2017A&A...600A..13A} derived the equations of $C_{\rm cf}$ and $R_{\rm phot}$ based on the empirical and synthetic spectral library, respectively.
\citet{2018A&A...619A..73L} and \citet{2023A&A...671A.162M} have provided estimates of $C_{\rm cf}$ and $R_{\rm phot}$ as functions of effective temperature.

The stellar surface flux now is relatively accurately determined in synthetic spectral model such as ATLAS, PHOENIX and MARCS \citep{2005A&A...442.1127M, 1995ApJ...445..433A, 2008A&A...486..951G}.
The synthetic spectral library PHOENIX was widely used in the calculation of chromospheric activity index based on the \ion{Ca}{II} H and K lines, e.g.,
\citet{2013A&A...549A.117M} estimated the stellar surface flux as a formula of $B-V$, and \citet{2023A&A...671A.162M} derived the relationship between $C_{\rm cf}$ and the stellar effective temperature.
\citet{2014MNRAS.445..270P} directly cross-matched each observed spectrum with the synthetic spectral library PHOENIX to derive an empirical chromospheric basal flux line.
In addition, the PHOENIX spectral library is also used to deduce the photospheric contribution (e.g., \citealt{2013A&A...549A.117M, 2017A&A...600A..13A, 2023A&A...671A.162M}).
\citet{2017A&A...600A..13A} pointed out that the photospheric contribution derived from the PHOENIX library is higher than that obtained from empirical spectra by \citet{1984ApJ...279..763N}.

With the development of large-scale photometric and spectroscopic surveys, statistical investigation of stellar chromosphere may disclose some novel phenomena \citep{2021Univ....7..440D}.
The Large Sky Area Multi-Object Fiber Spectroscopic Telescope (LAMOST, also named the Guoshoujing Telescope) has released massive spectral data since its pilot survey started in 2011 \citep{2012RAA....12.1197C, 2012RAA....12..723Z, 2012RAA....12.1243L}.
The spectra released by the Low-Resolution Spectroscopic Survey (LRS) of LAOMST cover the wavelength from 3700 to 9100\,{\AA} with a spectral resolving power ($R = \lambda/\Delta \lambda$) of about 1800 \citep{2012RAA....12..723Z}. 
A number of investigations of chromospheric activity have profited from the several spectral lines recorded by LAMOST, such as the \ion{Ca}{II} H and K lines, the ${\rm H}\alpha$ line and the \ion{Ca}{II} infrared
triplet (IRT) lines (e.g., \citealt{2019ApJ...887...84Z, 2020ApJS..247....9Z, 2021RAA....21..249B, 2023ApJS..264...12H, 2023Ap&SS.368...63H, 2024ApJS..272....6H}). In our previous work (\citealt{2022ApJS..263...12Z}, hereafter Paper I), we investigated the \ion{Ca}{II} H and K lines of LAMOST LRS spectra and provided a stellar chromospheric activity database, especially calibrated the $S$ index value of LAMOST to the scale of MWO. In this work, we dedicate to describe the chromospheric activity of solar-like stars based on the bolometric and photospheric calibrated indexes of LAMOST LRS spectra \citep{2021RNAAS...5....6H}.

This paper is organized as follows. In Section \ref{sec:data}, we describe the spectral data used in this work. In Section \ref{sec:process}, the detailed procedures of deriving the chromospheric activity indexes and their uncertainties are illustrated. In Section \ref{sec:results}, we present the database provided in this work and discuss the chromospheric activity based on the database. Finally we provide a brief summary and conclusion of this work in Section \ref{sec:conclusion}.

\section{Data Collection of Solar-like Stars} \label{sec:data}

We use the LAMOST LRS spectra in the Data Release 8 (DR8) v2.0\footnote{\url{http://www.lamost.org/dr8/v2.0/}}, which were observed between October 2011 and May 2020.
The LAMOST DR8 v2.0 comprises 10,633,515 LRS spectra, among which 6,684,413 spectra with determined stellar parameters have been published in the {\tt\string LAMOST LRS Stellar Parameter Catalog of A, F, G and K Stars} (hereafter referred to as the {\tt\string LAMOST LRS AFGK Catalog}). 
The stellar parameters such as effective temperature ($T_{\rm eff}$), surface gravity ($\log\,g$), metallicity ([Fe/H]), heliocentric radial velocity ($V_r$) and their corresponding uncertainties are afforded by the LAMOST Stellar Parameter Pipeline (LASP) \citep{2015RAA....15.1095L}.

We select the spectra of solar-like stars \citep{1996A&ARv...7..243C} by the effective temperatures around the Sun ($T_{\rm eff,\odot}  = 5777\,{\rm K}$, adopted as in \citealt{2012ApJ...752....5R}) that are in the range of $4800 \le T_{\rm eff} \le 6300\,{\rm K}$, and the metallicities around the Sun ([Fe/H]$_\odot=0.0\,{\rm dex}$) that are in the range of $-1.0<\text{[Fe/H]}<1.0\,{\rm dex}$.
The spectra of main-sequence stars are empirically separated from the giant sample by the criterion of $\log\,g$ as adopted in Paper I:
 \begin{equation} \label{eq:main-sequence_condition}
         \log\,g \ge 5.98-0.00035 \times T_{\rm eff}.
 \end{equation}
 
\begin{figure} 
	\resizebox{\hsize}{!}{\includegraphics{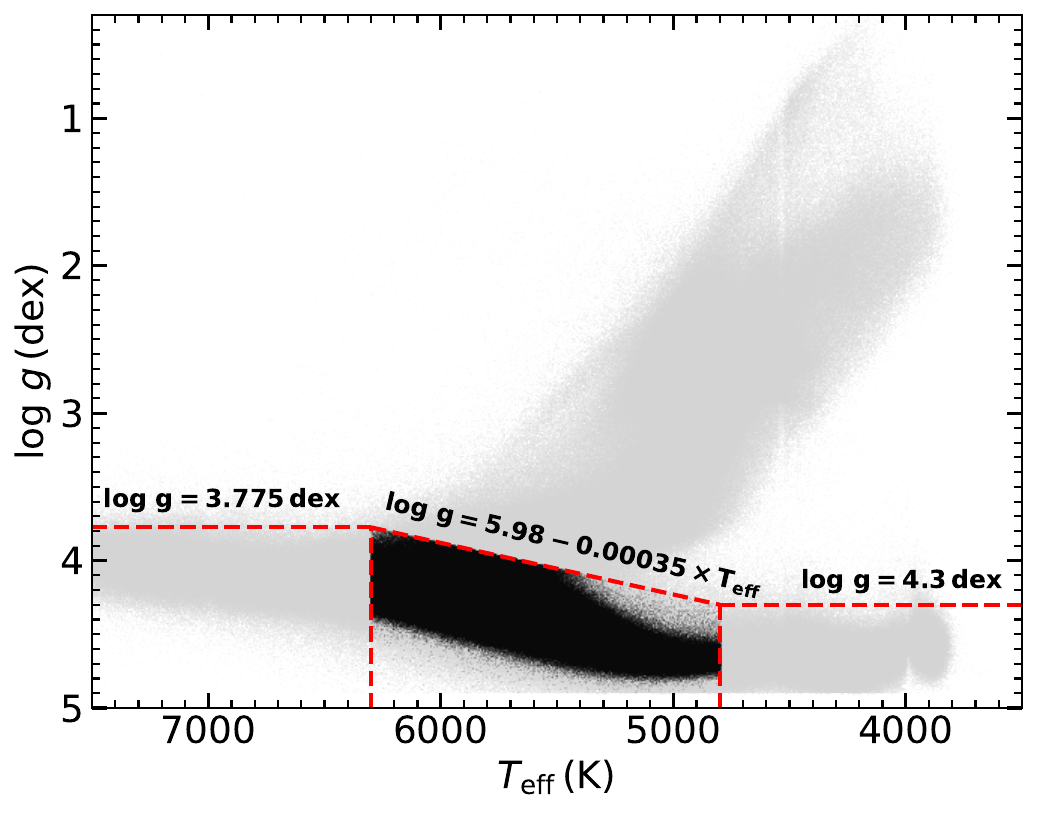}}
	\caption{The distribution of the selected spectral samples (black dots) of solar-like stars and the samples in {\tt\string LAMOST LRS AFGK Catalog} (gray dots) with $T_{\rm eff}$ in the range of 3500 to 7500 K. The vertical and horizontal red lines represent the empirical selection ranges of $T_{\rm eff}$ and $\log\,g$, respectively; the diagonal red line denotes the corresponding empirical formula in Equation \ref{eq:main-sequence_condition}. 
	}	
	\label{fig:sample_distribution}
\end{figure}

\begin{figure} 
	\resizebox{\hsize}{!}{\includegraphics{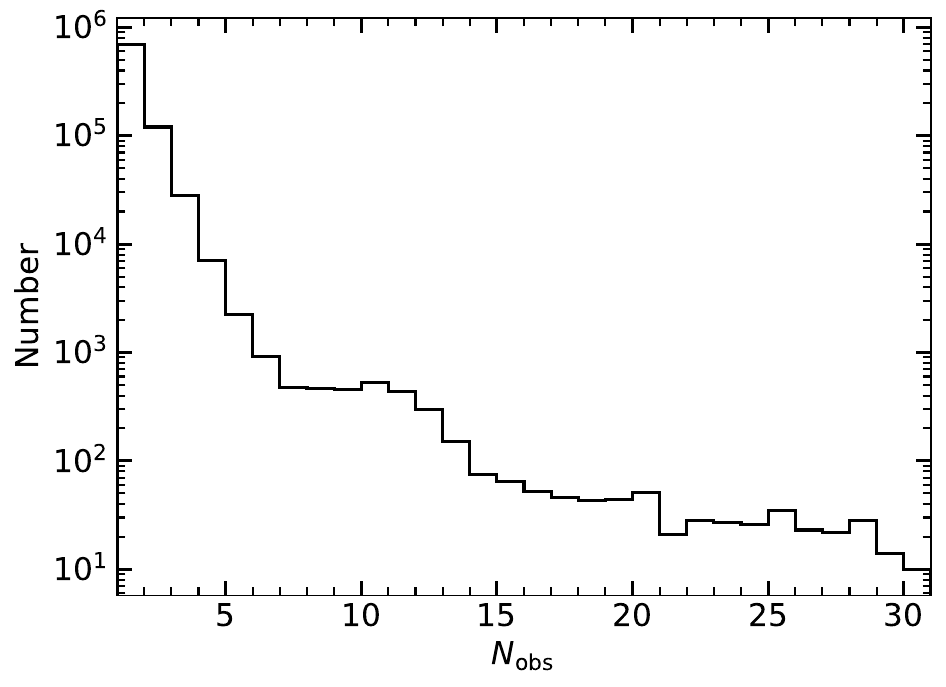}}
\caption{Histogram of the number of observations for the 861,505 solar-like stars.}
	\label{fig:Nobs_hist}
\end{figure}

The uncertainties of the chromospheric activity indexes are related to the uncertainties of the spectral fluxes and the corresponding stellar parameters derived from the LRS spectra which are predominantly impacted by the signal-to-noise ratio (S/N) parameters of LRS spectra.
The precision of spectral fluxes of \ion{Ca}{II} H and K lines and stellar parameters is primarily affected by the S/N in the $g$ and $r$ bands of LRS.
Therefore, the high-S/N spectra of solar-like stars are screened out by the S/N threshold ${\rm S/N}_g \ge 50.00$ and ${\rm S/N}_r \ge 71.43$ as adopted in Paper I.
A total of 1,149,216 spectra of solar-like stars are picked out from the {\tt\string LAMOST LRS AFGK Catalog}.
The band of \ion{Ca}{II} H and K lines used to derive the chromospheric activity index refer to the vacuum wavelength range of 3892.17-4012.20\,{\AA} (see Paper I).
The spectra in this band with zero or negative flux are discarded.
We eventually analyse the chromospheric activity based on 1,122,495 LAMOST LRS spectra of solar-like stars.
The distribution of the selected spectral samples is shown in Figure \ref{fig:sample_distribution}, where the gray dots represent the samples in {\tt\string LAMOST LRS AFGK Catalog}.
Since abundant stellar information is available in Gaia DR3 \citep{2023A&A...674A...1G}, we identified 861,505 solar-like stars in the selected spectra with {\tt\string gaia\_source\_id} obtainable in {\tt\string LAMOST LRS AFGK Catalog}.
Given the LAMOST is dedicated to the spectral survey of large sky areas, 81\% of the solar-like stars have been observed only once. 
Figure \ref{fig:Nobs_hist} displays the histogram of the observation numbers for 861,505 solar-like stars.

\begin{figure*} 
    \resizebox{\hsize}{!}{\includegraphics{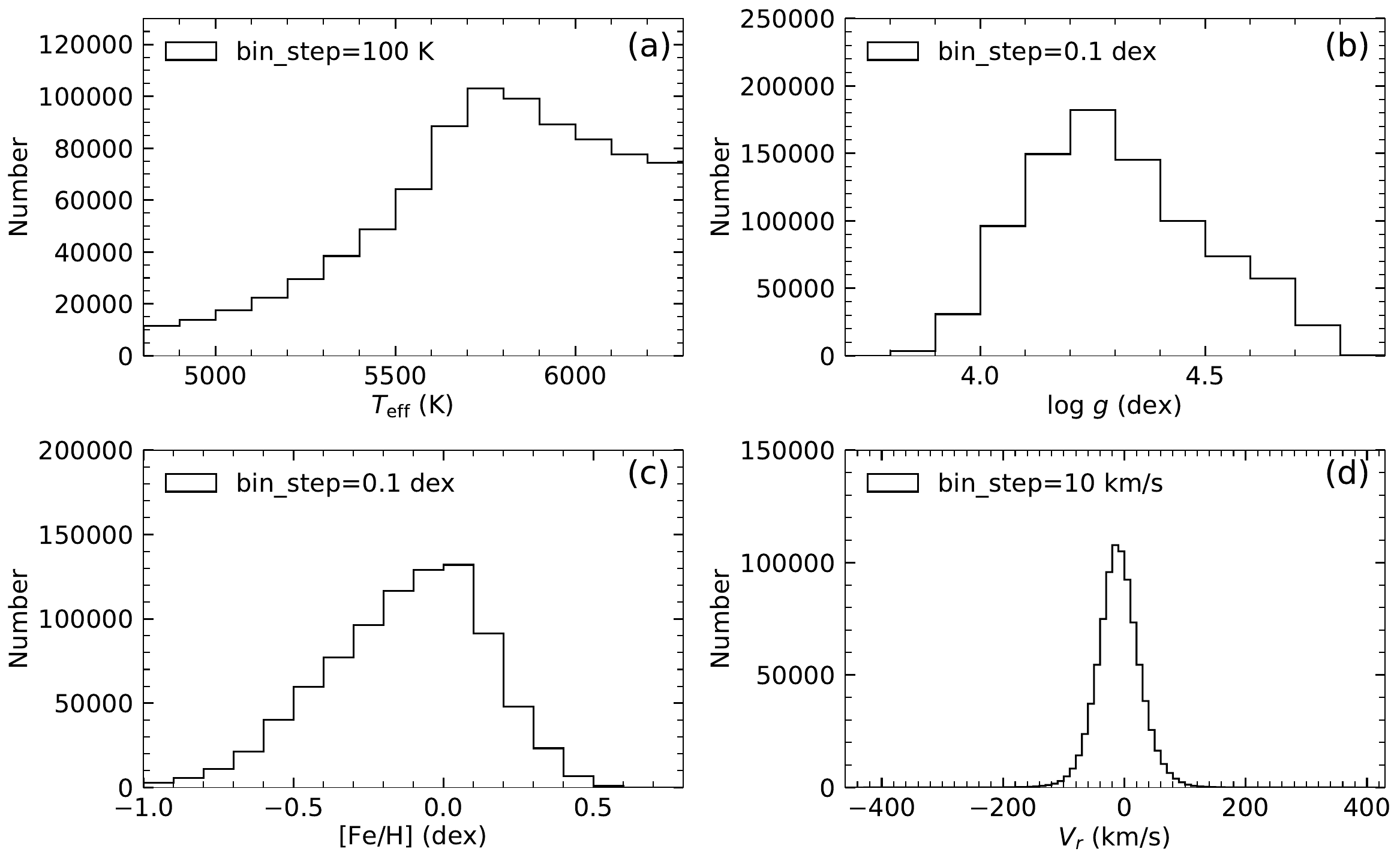}}
    \caption{Histograms of the (a) $T_{\rm eff}$, (b) $\log\,g$, (c) [Fe/H] and (d) $V_r$ of the 861,505 solar-like stars employed in this work.
    }
    \label{fig:teff-logg-feh-rv_hist}
\end{figure*}

In Figure \ref{fig:teff-logg-feh-rv_hist}, we show the histograms of the $T_{\rm eff}$, $\log\,g$, [Fe/H] and $V_r$ for the 861,505 solar-like stars used in this work.
If one star is observed more than once, the values of $T_{\rm eff}$, $\log\,g$, [Fe/H] and $V_r$ are taken as the median of the corresponding values of the multiple observed spectra.
As illustrated in the LAMOST DR8 release note \footnote{\url{https://www.lamost.org/dr8/v2.0/doc/release-note}}, the uncertainties of $T_{\rm eff}$, $\log\,g$, [Fe/H] and $V_r$ are relatively higher for S/N${_r}$ less than 30 and are relatively accurate for S/N${_r}$ greater than 50.
Based on our aforementioned S/N threshold of spectral selection (${\rm S/N}_g \ge 50.00$ and ${\rm S/N}_r \ge 71.43$), the  uncertainties of $T_{\rm eff}$, $\log\,g$, [Fe/H] and $V_r$ for the selected stars, which are provided by LASP, are approximately distributed around 25 K, 0.035 dex , 0.025 dex, and 3.5 ${\rm km\, s^{-1}}$, respectively.
The LAMOST DR8 release note also provides the parameter comparisons between LAMOST DR8 v2.0 LRS and DR16 \citep{2020ApJS..249....3A} of the Sloan Digital Sky Survey \citep{2000AJ....120.1579Y}.
Since the effective temperature plays an important role in Section \ref{sec:process}, we compare the $T_{\rm eff}$ provided by LASP with the results of \citet{2020MNRAS.499.3481A} in Appendix \ref{sec:acccuracy_SP}.

\section{Evaluation of the Chromospheric Activity Index of Solar-like Star}\label{sec:process}

The following steps are taken in the derivation of the chromospheric activity index $R'_{\rm HK}$ of solar-like stars based on the \ion{Ca}{II} H and K lines of LAMOST LRS: 
(1) definition of $S_{\rm MWO}$ for LAMOST LRS,
(2) conversion of $S_{\rm MWO}$ to $R_{\rm HK}$,
(3) derivation of the bolometric and photospheric calibrated chromospheric activity index $R'_{\rm HK}$,
(4) estimation of the uncertainty of chromospheric activity indexes.

\subsection{Definition of \texorpdfstring{$S_{\rm MWO}$}{S\_MWO} for LAMOST LRS}\label{sec:Sindex}
The stellar chromospheric activity index has been widely studied and broadened based on the emission of the \ion{Ca}{II} H and K lines (e.g., \citealt{1968ApJ...153..221W, 1978PASP...90..267V, 1991ApJS...76..383D, 1995ApJ...438..269B, 2007AJ....133..862H, 2017ApJ...835...25E, 2018A&A...616A.108B}).
In 1966, the two-channel HKP-1 spectrophotometer was employed in Mount Wilson Observation to monitor the emission of stellar \ion{Ca}{II} H and K lines \citep{1968ApJ...153..221W, 2007AJ....133..862H}.
One channel was used to collect data in the 25\,{\AA} rectangular bands located in the red and violet sides of the \ion{Ca}{II} H and K lines.
The counts in the reference bands of this channel were noted as $N_{\mathcal{RV}}$.
The other channel was used to measure the emission in the 1\,{\AA} rectangular bands centered at \ion{Ca}{II} H or K lines.
After completing the counts in either H or K lines, the relative instrumental fluxes $F_{\mathcal{H}} = N_{\mathcal{H}}/N_{\mathcal{RV}}$ or $F_{\mathcal{K}} = N_{\mathcal{K}}/N_{\mathcal{RV}}$ could be collected, where $N_{\mathcal{H}}$ and $N_{\mathcal{K}}$ are the counts in the 1\,{\AA} rectangular bands centered at \ion{Ca}{II} H and K lines.
\citet{1968ApJ...153..221W} employed $F = \frac{1}{2}(F_{\mathcal{H}}+F_{\mathcal{K}})$ to assess the emission of \ion{Ca}{II} H and K lines collected by the HKP-1.

In view of the instrumental effects and certain limitations in the HKP-1,
\citet{1978PASP...90..267V} introduced the HKP-2, a four-channel spectrophotometer in 1977.
The H and K channels collected the two 1.09\,{\AA} full width at half maximum (FWHM) triangular bandpasses in the line cores of \ion{Ca}{II} H and K lines centered in the air wavelengths of 3968.47 and 3933.66\,{\AA}, respectively. 
In addition, the R and V channels measured the two 20\,{\AA} rectangular bandpasses on the red and violet sides of the \ion{Ca}{II} H and K lines (wavelength ranges in air being 3991.07--4011.07\,{\AA} and 3891.07--3911.07\,{\AA}, respectively). 
The H, K, R and V channels are exposed sequentially and rapidly, with the exposure time of the H and K channels being eight times that of the R and V channels.
To align the HKP-2 data with the HKP-1 data, \citet{1978PASP...90..267V} performed a calibration of the HKP-2 data to match the HKP-1 data by
\begin{equation} \label{eq:S_MWO-HKP}
    S_{\rm MWO} = \alpha \cdot \frac{N_{\rm H}+N_{\rm K}}{N_{\rm R}+N_{\rm V}},
\end{equation}
where $N_{\rm H}$, $N_{\rm K}$, $N_{\rm R}$ and $N_{\rm V}$ are the counts in the H, K, R and V channels of HKP-2, respectively, and the scaling factor $\alpha=2.4$ is applied to ensure the consistency of the results between HKP-2 and HKP-1 \citep{1978PASP...90..267V, 1991ApJS...76..383D}.

Previous studies have utilized various bandpasses and definitions of chromospheric $S$ index to calibrate their measurements from different instruments to the scale of MWO \citep{2003AJ....126.2048G, 2004ApJS..152..261W, 2011A&A...531A...8J, 2022ApJ...929...80B}.
We discussed two typically definitions of the $S$ index in Paper I, namely $S_{\rm rec}$ and $S_{\rm tri}$, which are computed from the \ion{Ca}{II} H and K lines using 1\,{\AA} rectangular bandpasses and 1.09\,{\AA} FWHM triangular bandpasses, respectively.
As a conclusion, these two kinds of definition of $S$ index is comparable for investigating the stellar chromospheric activity based on the \ion{Ca}{II} H and K lines observed by LAMOST LRS.
The $S_{\rm tri}$ is defined as
\begin{equation} \label{eq:S_tri}
    S_{\rm tri} = \frac{\widetilde{H}_{\rm tri} + \widetilde{K}_{\rm tri}}{\widetilde{R}+\widetilde{V}},
\end{equation}
where $\widetilde{R}$ and $\widetilde{V}$ represent the mean fluxes in the 20\,{\AA} rectangular bandpasses centered in the vacuum wavelength of 4002.20 and 3902.17\,{\AA},
$\widetilde{H}_{\rm tri}$ and $\widetilde{K}_{\rm tri}$ are the mean fluxes in the 1.09\,{\AA} FWHM triangular bandpasses centered in the vacuum wavelength of 3969.59 and 3934.78\,{\AA} \citep{2011arXiv1107.5325L, 2022ApJS..263...12Z}.
Since the vacuum wavelength is adopted in LAOMST LRS spectra, the above vacuum wavelength values of the bandpasses center are converted from the corresponding wavelength values in air, see Paper I. 
The relationship between the value of vacuum wavelength and air wavelength is obtained from \citet{1996ApOpt..35.1566C}.

A denser and uniform distribution of wavelength is instrumental in integration of spectral flux. 
To estimate the mean fluxes in each bandpass, the step of spectral wavelength were linearly interpolated to 0.001\,{\AA}.
The wavelength shift caused by radial velocity could not be ignored, because the bandpasses used for calculating are narrow.
We calibrate the spectral wavelength to the values in the rest frame before the calculation of chromospheric activity index.
The pretreatment of wavelength based on $V_r$ is introduced in Paper I.

In Paper I, 65 common stars were identified by cross-matching the database in that work and the $S_{\rm MWO}$ catalog of MWO in \citet{1991ApJS...76..383D}.
A relationship between the $S$ indexes of LAMOST and the $S_{\rm MWO}$ was introduced to calibrate the result of LAMOST to the scale of MWO.
The relationship between the $S_{\rm tri}$ and $S_{\rm MWO}$ can be expressed by an exponential formula 
\begin{equation} \label{eq:S_tri_vs_S_MWO}
    S_{\rm MWO} = \textrm{\large e}^{\,6.913\,S_{\rm tri}-3.348},
\end{equation}
and the detailed technological process can be found in Paper I.

\begin{figure} 
	\resizebox{\hsize}{!}{\includegraphics{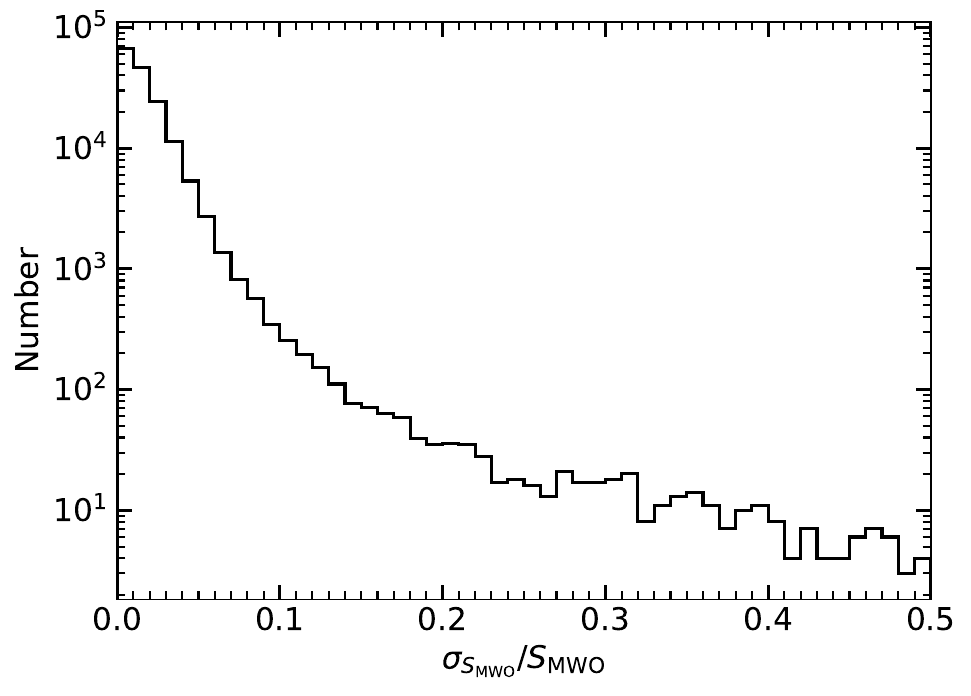}}
	\caption{Histogram of  $\sigma_{S_{\rm MWO}}/S_{\rm MWO}$ for stars with more than one observation, where the $\sigma_{S_{\rm MWO}}$ is the standard deviation of $S_{\rm MWO}$.
	}
	\label{fig:S_MWO_std}
\end{figure}

The histogram of $\sigma_{S_{\rm MWO}}/S_{\rm MWO}$ for stars with more than one observation is shown in Figure \ref{fig:S_MWO_std}, where the $\sigma_{S_{\rm MWO}}$ represent the standard deviation of $S_{\rm MWO}$. For the majority of the samples (98.98\%), the values of $\sigma_{S_{\rm MWO}}/S_{\rm MWO}$ are less than 0.1.

\subsection{Conversion of \texorpdfstring{$S_{\rm MWO}$}{S\_MWO} to \texorpdfstring{$R_{\rm HK}$}{R\_HK}}\label{sec:RHK}

In order to connect with physical quantity, the $S_{\rm MWO}$ can be described by the stellar surface fluxes as
\begin{equation} \label{eq:SMWO_vs_F}
    S_{\rm MWO} = 8\alpha \cdot \frac{\mathcal{F}_{\rm HK}}{\mathcal{F}_{\rm RV}},
\end{equation}
where $\mathcal{F}_{\rm HK}$ is the stellar surface flux in the 1.09\,{\AA} FWHM H and K bands, and $\mathcal{F}_{\rm RV}$ represents the stellar surface flux in the 20\,{\AA} R and V bands \citep{1978PASP...90..267V,1982A&A...107...31M, 1984A&A...130..353R, 2007AJ....133..862H, 2013A&A...549A.117M,2023A&A...671A.162M}.
The constant 8 comes from the aforementioned different exposure times in HKP-2, and $\alpha$ represents the scaling factor in Equation \ref{eq:S_MWO-HKP} \citep{1978PASP...90..267V}.
The ${\mathcal{F}_{\rm RV}}$ mainly depends on the stellar atmospheric parameters, thus can be derived from empirical spectral library (e.g., \citealt{1982A&A...107...31M, 1984A&A...130..353R, 2007A&A...469..309C, 2015MNRAS.452.2745S, 2017A&A...600A..13A, 2018A&A...619A..73L}) or synthetic spectral library (e.g., \citealt{2013A&A...549A.117M, 2014MNRAS.445..270P, 2023A&A...671A.162M}).
By combining the $S_{\rm MWO}$ with corresponding stellar continuum spectra of the empirical or synthetic spectral library, a new active index $R_{\rm HK}$ can be constructed as follows, which describes the emission of \ion{Ca}{II} H and K lines more physically than $S_{\rm MWO}$ (e.g., \citealt{1979ApJS...41...47L, 1982A&A...107...31M, 1984A&A...130..353R, 1984ApJ...279..763N}).

The stellar surface flux $\mathcal{F}_{\rm HK}$ in the 1.09\,{\AA} FWHM H and K bands can be normalised by the bolometric flux as
\begin{equation} \label{eq:RHK_vs_FHK}
    R_{\rm HK} = \frac{\mathcal{F}_{\rm HK}}{\sigma T_{\rm eff}^4},
\end{equation}
where $T_{\rm eff}$ is the stellar effective temperature and $\sigma = 5.67  \times 10^{-5}\,{\rm erg}\,{\rm cm}^{-2}\,{\rm s}^{-1}\,{\rm K}^{-4}$ is the Stefan-Boltzmann constant.
$R_{\rm HK}$ is not related to the continuum flux around the \ion{Ca}{II} H and K lines which is governed by the stellar effective temperature, thus it can be used to compare stars with different spectral types.

A widely used form of the relationship between $S_{\rm MWO}$ and $R_{\rm HK}$ can be expressed as
\begin{equation} \label{eq:RHK_vs_Ccf}
    R_{\rm HK} = K \cdot \sigma^{-1} \cdot 10^{-14} \cdot C_{\rm cf} \cdot S_{\rm MWO},
\end{equation}
where $C_{\rm cf}$ is the bolometric factor, and $10^{-14}$ is an arbitrary factor \citep{1982A&A...107...31M, 1984A&A...130..353R}.
The factor $K$ is in unit of ${\rm erg}\ {\rm cm}^{-2}\ {\rm s}^{-1}$, 
which was introduced by \citet{1982A&A...107...31M} to convert the relative flux $F_{\rm HK}$ (in arbitrary units) 
 \begin{equation} \label{eq:FHK_vs_Ccf}
    F_{\rm HK} = C_{\rm cf} \cdot T_{\rm eff}^4 \cdot 10^{-14} \cdot S_{\rm MWO}
\end{equation}
to stellar surface flux $\mathcal{F}_{\rm HK}$ by 
\begin{equation} \label{eq:F_HK_vs_K}
    \mathcal{F}_{\rm HK} = K \cdot F_{\rm HK}.
\end{equation}

The factor $C_{\rm cf}$ in Equation \ref{eq:RHK_vs_Ccf} was derived by the pioneer studies of \citet{1982A&A...107...31M} and \citet{1984A&A...130..353R} (see e.g., \citealt{1984ApJ...279..763N, 1996AJ....111..439H, 2004ApJS..152..261W, 2007AJ....133..862H, 2017ApJ...835...25E, 2019MNRAS.485.5096K, 2021ApJ...914...21S,  2021A&A...646A..77G}).
\citet{1982A&A...107...31M} first introduced and deduced the $C_{\rm cf}$ as a function of $B-V$ for main-sequence stars with $0.45 \le B-V \le 1.5$.
Subsequently, \citet{1984A&A...130..353R} broaden the $C_{\rm cf}$ to $ 0.3 \le B-V \le 1.6$ for FGK type main-sequence stars.
To describe the chromospheric activity of M dwarf, \citet{2007A&A...469..309C} calibrated $C_{\rm cf}$ to the range $0.45 \le B-V \le 1.81$ for stars with spectral type ranging from F6 to M5, and \citet{2015MNRAS.452.2745S} calibrated the $C_{\rm cf}$ in the range of $0.4 \lesssim B-V \lesssim 1.9$. 
On the other hand, \citet{2017A&A...600A..13A} derived the $C_{\rm cf}$ in the range of $0.54 \le B-V \le 1.90$ to include the M dwarf, and they prefer the forms of $C_{\rm cf}$ described by the color indexes $I-K$ and $V-K$.
\citet{2018A&A...619A..73L} calibrated the $C_{\rm cf}$ to a function of $T_{\rm eff}$.

The preceding researches of $C_{\rm cf}$ are based on empirical method, which can also be obtained from synthetic spectral library.
The PHOENIX model atmospheres were utilized by \citet{2013A&A...549A.117M} to obtain $\mathcal{F}_{\rm RV}$ for main-sequence stars, and the relation between $\mathcal{F}_{\rm RV}$ and $B-V$ is given as
\begin{equation}\label{eq:F_RV_2013}
    {\rm log} \,\frac{\mathcal{F}_{\rm RV}}{19.2} =  8.33 - 1.79(B - V),
\end{equation}
where $0.44 \le B-V \le 1.6$, and the constant 19.2 is equal to the scaling factor $8\alpha$ in Equation \ref{eq:SMWO_vs_F}.
 $\mathcal{F}_{\rm RV}$ is comparable with $C_{\rm cf}$, which can be transfer to $C_{\rm cf}$ through 
\begin{equation}\label{eq:Ccf_vs_F_RV}
    C_{\rm cf} = \frac{\mathcal{F}_{\rm RV}}{8\alpha \cdot K \cdot T^4_{\rm eff} \cdot 10^{-14}},
\end{equation}
based on Equations \ref{eq:SMWO_vs_F}, \ref{eq:FHK_vs_Ccf} and \ref{eq:F_HK_vs_K}.
Through matching the PHOENIX spectral library with observed spectra, 
\citet{2014MNRAS.445..270P} introduced a quadratic formula of $C_{\rm cf}$ within  $0.44 \le B-V \le 1.33$ for luminosity classes V and IV with $\log\,g$ between 5.0 and $3.5\,{\rm dex}$.
\citet{2023A&A...671A.162M} derived the $C_{\rm cf}$ as a fifth-order function of $T_{\rm eff}$ based on the synthetic spectral library of PHOENIX.

The $C_{\rm cf}$ in different researches described above 
generally can be expressed with a polynomial
\begin{equation}\label{eq:log_Ccf_vs_X}
    {\rm log} \ C_{\rm cf} (X) = \sum_{i=0}^{5} C_{i}X^{i} ,
\end{equation}
where $X$ represents $B-V$ or $T_{\rm eff}$, and $C_{i}$ (i=0,...,5) are the corresponding coefficients which are presented in Table \ref{tab:C_cf}.

\begin{table*} 
\caption{Coefficients for the ${\rm log} \ C_{\rm cf}$ expression in Equation \ref{eq:log_Ccf_vs_X} in different researches. \label{tab:C_cf}}
\begin{tabular}{cccccccc}
\toprule\toprule
Source & $C_0$ & $C_1$ & $C_2$ & $C_3$ & $C_4$ & $C_5$  & Valid Range  \\
  &   &   &   &   &   &    & of $B-V$ or $T_{\rm eff}$  \\
\hline
(1) & -0.47 & 2.84 & -3.91 & 1.13 & / & /  & $0.45\le B-V \le 1.5$ \\
(2) & 0.24 & 0.43 & -1.33 & 0.25 & / & / & $0.3 \le B-V \le 1.6$ \\ 
(3) & 0.8 & -1.41 & 0.55 & -0.33 & / & / & $0.45 \le B-V \le 1.81$ \\
(4) & 0.66 & -1.11 & / & / & / & / & $0.44 \le B-V \le 1.33$ \\
(5) & 0.668 & -1.270 & 0.645 & -0.443 & / & / & $0.4 \lesssim B-V \lesssim 1.9$ \\ 
(6) & 0.669 & -0.972 & 0.109 & -0.203 & / & / & $0.54 \le B-V \le 1.90$ \\
(7) & -7.31 & $2.25 \times 10^{-3}$ & $-1.70 \times 10^{-7}$ & / & / & / & $4350 \le T_{\rm eff} \le 6500\,{\rm K}$ \\
(8) & -29.679 & 0.026864 & $-1.0268 \times 10^{-5}$ & $1.9866 \times 10^{-9}$ & $-1.9017 \times 10^{-13}$ & $7.1548 \times 10^{-18}$ & $2300 \le T_{\rm eff} \le 7200\,{\rm K}$ \\
\hline
\end{tabular}
\tablebib{
    (1)~\citet{1982A\string&A...107...31M};
    (2)~\citet{1984A\string&A...130..353R} (used for the calculation of $R_{\rm HK,classic}$);
    (3)~\citet{2007A\string&A...469..309C};
    (4)~\citet{2014MNRAS.445..270P};
    (5)~\citet{2015MNRAS.452.2745S}；
    (6)~\citet{2017A\string&A...600A..13A}；
    (7)~\citet{2018A\string&A...619A..73L}；
    (8)~\citet{2023A\string&A...671A.162M}.}
\end{table*}

The $S_{\rm MWO}$ can be derived from observed spectra, and $C_{\rm cf}$ can be estimated by stellar color index or $T_{\rm eff}$ as described above.
The remaining coefficient in Equation \ref{eq:RHK_vs_Ccf} to be determined is the factor $K$.
\citet{1982A&A...107...31M} deduced the $K = (0.76 \pm 0.11) \times 10^{6}\ {\rm erg}\ {\rm cm}^{-2}\ {\rm s}^{-1}$ based on the investigation of \citet{1979ApJS...41...47L}, thus $K \cdot \sigma^{-1} \cdot 10^{-14} = 1.34 \times 10^{-4}$ , which is frequently adopted in the relevant works (e.g., \citealt{ 2007A&A...469..309C, 2014MNRAS.445..270P, 2015MNRAS.452.2745S, 2018A&A...616A.108B, 2021A&A...646A..77G}).
\citet{1984A&A...130..353R} derived $K = (1.29 \pm 0.19) \times 10^{6}\ {\rm erg}\ {\rm cm}^{-2}\ {\rm s}^{-1}$ based on the solar $S$ index $S_{\rm MWO,\sun}=0.160$ \citep{1983A&A...124...43O} and the color index of the Sun $(B-V)_\sun = 0.665$ \citep{1980A&A....91..221H}.
Additionally, \citet{2007AJ....133..862H} conducted a recalibration of the $K$ value, obtaining a result of $(1.07 \pm 0.13) \times 10^{6}\ {\rm erg}\ {\rm cm}^{-2}\ {\rm s}^{-1}$. They pointed out that the discrepancy between their result and the $K$ value reported by \citet{1984A&A...130..353R} is mainly due to the adoption of a different solar $B-V$ value, which they took to be 0.642 \citep{1996A&ARv...7..243C}.
$K \cdot \sigma^{-1} \cdot 10^{-14} = 1.887 \times 10^{-4}$ is gradually adopted in recent works (e.g., \citealt{2017A&A...600A..13A, 2020AJ....160..269M, 2022ApJ...929...80B, 2023A&A...671A.162M}).

In this work, the value of $R_{\rm HK}$ derived from the method in the classic literature for the LRS spectra is denoted as $R_{\rm HK,classic}$, which is calculated by utilizing the $C_{\rm cf}$ from \citet{1984A&A...130..353R} (row 2 in Table \ref{tab:C_cf}) and $K = 0.76 \times 10^{6}\,{\rm erg}\,{\rm cm}^{-2}\,{\rm s}^{-1}$ from \citet{1982A&A...107...31M}.
Since the value of $B-V$ is needed for the estimation of $R_{\rm HK,classic}$, we use the relation between $T_{\rm eff}$ and $B-V$ introduced in \citet{1984ApJ...279..763N} to transform $T_{\rm eff}$ to $B-V$ when calculating $R_{\rm HK,classic}$, which is based on the research of \citet{1966ARA&A...4..193J}.
The transformation is given by
\begin{equation} \label{eq:teff_vs_bv}
\log\,T_{\rm eff} = 3.908 - 0.234(B-V),
\end{equation}
 in the range $0.4 < B-V < 1.4$.

Based on Equations \ref{eq:SMWO_vs_F} and \ref{eq:RHK_vs_FHK}, we can express $R_{\rm HK}$ by $S_{\rm MWO}$ and $\mathcal{F}_{\rm RV}$ as
\begin{equation} \label{eq:RHK_vs_F_RV}
    R_{\rm HK} = \frac{ S_{\rm MWO} \cdot \mathcal{F}_{\rm RV}}{8\alpha} \cdot \frac{1}{\sigma T_{\rm eff}^4}.
\end{equation}
As described above, recent studies have demonstrated that the PHOENIX model is a useful tool for deriving the stellar surface flux $\mathcal{F}_{\rm RV}$.
In this work, besides $R_{\rm HK,classic}$ we also utilize the spectral library of PHOENIX to estimate $\mathcal{F}_{\rm RV}$, and then derive $R_{\rm HK}$ through Equation \ref{eq:RHK_vs_F_RV}, denoted as $R_{\rm HK,PHOENIX}$.
Because the detailed stellar atmospheric parameters ($T_{\rm eff}$, $\log\,g$ and [Fe/H]) are available for LAMOST (see Section \ref{sec:data}), and the $\mathcal{F}_{\rm RV}$ values estimated from the PHOENIX synthetic spectra library are related to these stellar parameters, we evaluate the values of $R_{\rm HK,PHOENIX}$ taking these parameters into account.

\begin{figure} 
    \resizebox{\hsize}{!}{\includegraphics{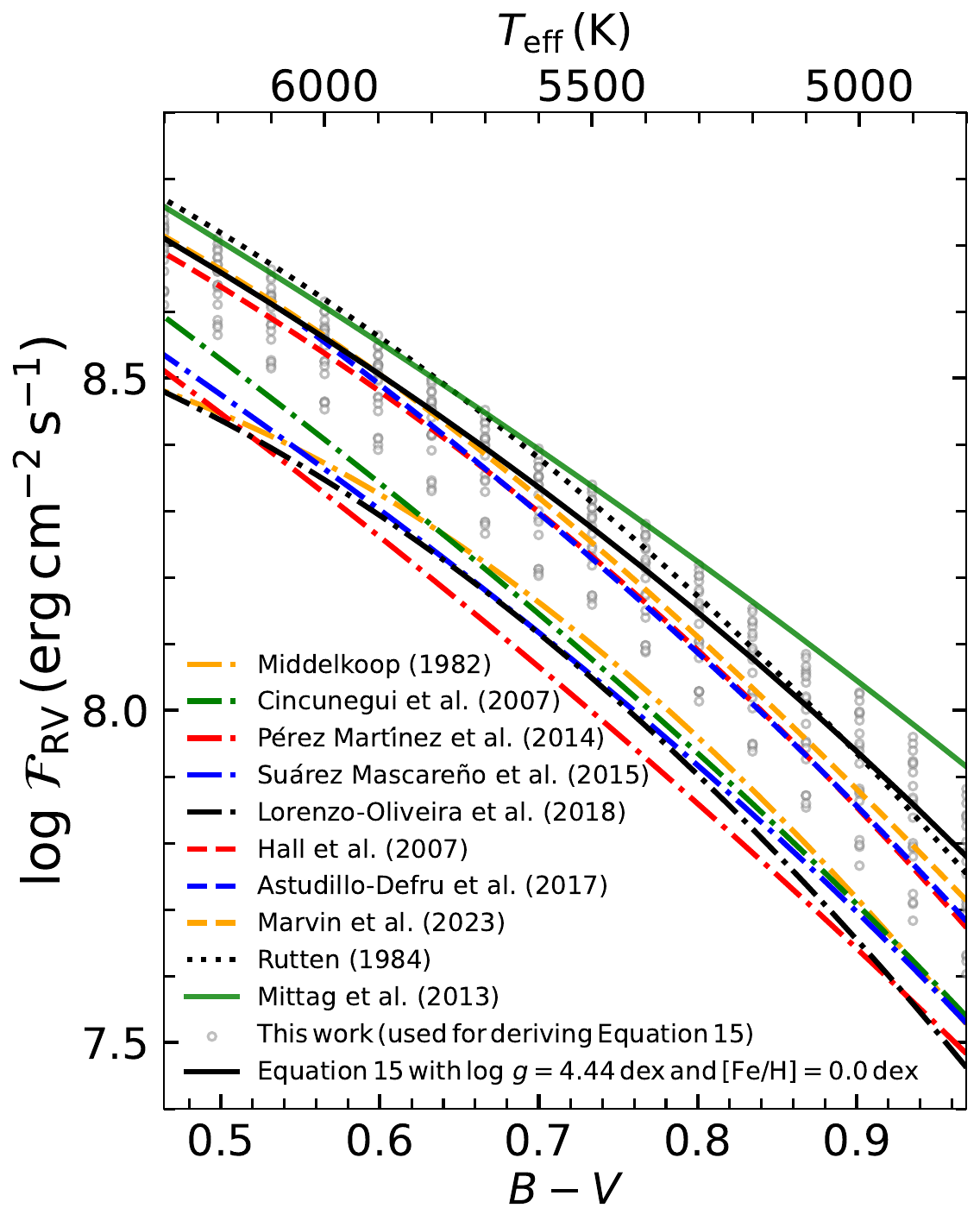}}
    \caption{Relations between $\log\, \mathcal{F}_{\rm RV}$ and $B-V$ (or $T_{\rm eff}$) adopted in different researches. For the works using $C_{\rm cf}$ and $K$ instead of $\mathcal{F}_{\rm RV}$, the values of $\mathcal{F}_{\rm RV}$ are estimated by Equation \ref{eq:Ccf_vs_F_RV}.
    The dashed-dotted, dashed and dotted lines represent the results using $K = 0.76 \times 10^{6}$, $1.07 \times 10^{6}$ and $ 1.29 \times 10^{6} \,{\rm erg}\,{\rm cm}^{-2}\,{\rm s}^{-1}$, respectively.
    The gray circles represent the $\mathcal{F}_{\rm RV}$ values estimated from the PHOENIX model used for deriving Equation \ref{eq:F_RV_vs_teff-logg-feh}.
    The black solid line is derived from Equation \ref{eq:F_RV_vs_teff-logg-feh} with $\log\,g = 4.44\,{\rm dex}$ and ${\rm [Fe/H]}=0.0\,{\rm dex}$.
    }
    \label{fig:Frv-teff}
\end{figure}

\citet{2013A&A...553A...6H} published a high-resolution synthetic spectral library\footnote{\url{http://phoenix.astro.physik.uni-goettingen.de/}} based on the version 16 of the PHOENIX model atmospheres.
The stellar atmospheric parameter space of their library covers $2300 \le T_{\rm eff} \le 12000\,{\rm K}$, $0.0 \le \log\,g \le 6.0\,{\rm dex}$ and $-4.0 \le {\rm [Fe/H]} \le 1.0\,{\rm dex}$. 
In \citet{2021A&A...649A..97L}, a comparison between the PHOENIX synthetic spectra library and empirical spectra was conducted, and their results show that the spectra of \citet{2013A&A...553A...6H} exhibit good consistency with the empirical spectra in the effective temperature range down to about $4000\,{\rm K}$.
Considering the stellar parameters space of the LAMOST LRS spectra of solar-like stars used in this work as described in Section \ref{sec:data}, we utilize the spectra in \citet{2013A&A...553A...6H} within the range of $4800 \le T_{\rm eff} \le 6300\,{\rm K}$, $3.5 \le \log\,g \le 5.0\,{\rm dex}$ and $-1.0 \le {\rm [Fe/H]} \le 1.0\,{\rm dex}$.
A total of 320 high-resolution synthetic spectra in this parameter range are collected to calculate the value of $\mathcal{F}_{\rm RV}$.
We fitted $\log\,\mathcal{F}_{\rm RV}$ by a ternary quadratic polynomial 
\begin{equation}\label{eq:F_RV_vs_teff-logg-feh}
\begin{aligned}
    {\rm log} \ \mathcal{F}_{\rm RV} 
    = &-138.7639
    + 70.3122 X
    + 0.3893 Y
    -2.3216 Z, \\
    &-0.0806 X  \cdot Y 
     -0.5840 X  \cdot Z
    + 0.0124 Y  \cdot Z \\
     &-8.2986 X^{2}
    -0.01242 Y^{2} 
    -0.0351 Z^{2},
\end{aligned}
\end{equation}
where the $X$, $Y$ and $Z$ represent $\log\,T_{\rm eff}$, $\log\,g$ and [Fe/H], respectively.
The fitting coefficients are calculated by the nonlinear least square method through the python module {\tt\string curve\_fit} of {\tt\string scipy.optimize} \citep{2020NatMe..17..261V}.
In Figure \ref{fig:Frv-teff}, we present the relationships between $\mathcal{F}_{\rm RV}$ and $B-V$ (or $T_{\rm eff}$) adopted in different researches.
The $\mathcal{F}_{\rm RV}$ values of the PHOENIX spectra used for deriving Equation \ref{eq:F_RV_vs_teff-logg-feh} are exhibited with the gray circles.
The black solid line in Figure \ref{fig:Frv-teff} is derived from Equation \ref{eq:F_RV_vs_teff-logg-feh} with $\log\,g = 4.44\,{\rm dex}$ and ${\rm [Fe/H]}=0.0\,{\rm dex}$ (solar parameters).
It can be seen from Figure \ref{fig:Frv-teff} that the results of \citet{2007AJ....133..862H} ($K = 1.07 \times 10^{6}\,{\rm erg}\,{\rm cm}^{-2}\,{\rm s}^{-1}$ and $C_{\rm cf}$ taken from \citet{1982A&A...107...31M}) using empirical spectra library is relatively close to the results calculated from the PHOENIX synthetic spectral library.

\begin{figure} 
    \resizebox{\hsize}{!}{\includegraphics{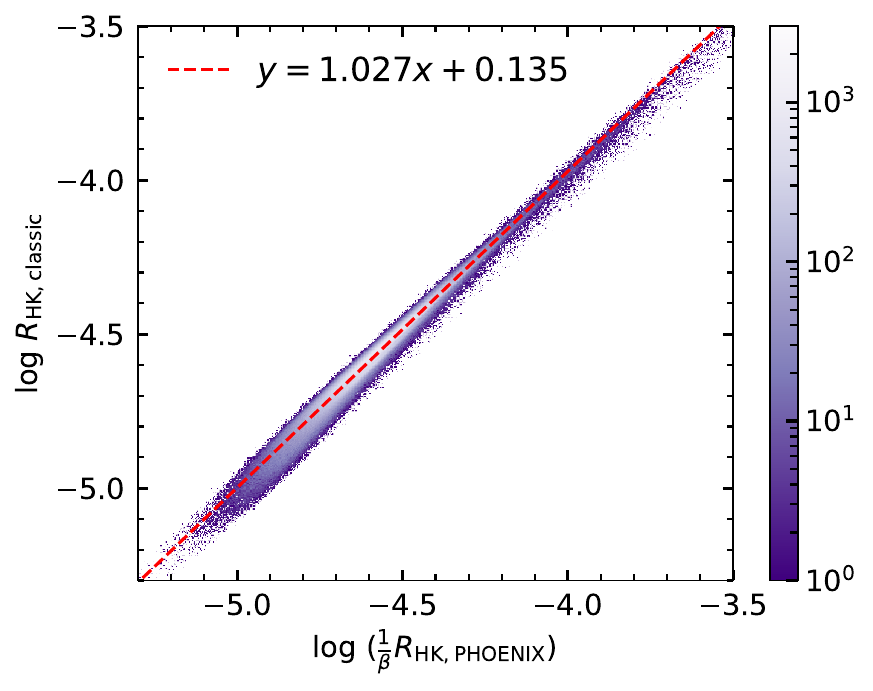}}
    \caption{Scatter diagram between $\log\,(\frac{1}{\beta}R_{\rm HK,PHOENIX})$ and $\log\,R_{\rm HK,classic}$ for the LAMOST LRS spectra of solar-like stars used in this work.
    The color scale indicates the density of data points.
	The dashed line represents the linear fit between  $\log\,(\frac{1}{\beta}R_{\rm HK,PHOENIX})$ and $\log\,R_{\rm HK,classic}$.
	}
    \label{fig:R_HK_PHOENIX-R_HK_classic}
\end{figure}

The $\mathcal{F}_{\rm HK,PHOENIX}$ value of the Sun is estimated to be $(2.423 \pm 0.007) \times 10^{6}\,{\rm erg}\,{\rm cm}^{-2}\,{\rm s}^{-1}$, which is calculated by Equations \ref{eq:SMWO_vs_F} and \ref{eq:F_RV_vs_teff-logg-feh} with $T_{\rm eff} = 5777\,{\rm K}$, $\log\,g=4.44\,{\rm dex}$ , ${\rm [Fe/H]} = 0.0\,{\rm dex}$ and $S_{\rm MWO,\odot}=0.1694 \pm 0.0005$. The selected values of solar $T_{\rm eff}$ and $\log\,g$ are following \citet{2012ApJ...752....5R}. The $S_{\rm MWO,\odot}$ is the mean value of the solar $S$ index which is obtained from the MWO HKP-2 measurements by \citet{2017ApJ...835...25E}.
The solar $\mathcal{F}_{\rm HK}$ values estimated by \citet{1983A&A...124...43O}, \citet{2007AJ....133..862H} and \citet{2013A&A...549A.117M} are $(2.17 \pm 0.32) \times 10^{6}$, $(2.12 \pm 0.25) \times 10^{6}$ and $(2.47 \pm 0.10) \times 10^{6}\,{\rm erg}\,{\rm cm}^{-2}\,{\rm s}^{-1}$, respectively.
Our evaluation of $\mathcal{F}_{\rm HK,\odot}$ value is consistent with those values estimated in previous investigations with a slight deviation. The deviation may originate from the different spectral fluxes in the PHOENIX model, the different choices of solar atmospheric parameters, and the different value of $S_{\rm MWO,\odot}$.
As a result, the values of $R_{\rm HK,PHOENIX}$ are relatively higher than $R_{\rm HK,classic}$, with a boost factor $\beta = 1.6$.
Figure \ref{fig:R_HK_PHOENIX-R_HK_classic} displays a comparison between the values of $\log\,(\frac{1}{\beta}R_{\rm HK,PHOENIX})$ and $\log\,R_{\rm HK,classic}$ for the LAMOST LRS spectra of solar-like stars used in this work. 
The correlation between them can be fitted by a linear formula
\begin{equation}\label{eq:R_HK_PHOENIX_vs_R_HK_classic}
    \log\,R_{\rm HK,classic} = 1.027\log\,(\frac{1}{\beta}R_{\rm HK,PHOENIX})+0.135.
\end{equation}

\subsection{Derivation of the Bolometric and Photospheric Calibrated Chromospheric Activity Index \texorpdfstring{$R'_{\rm HK}$}{Rp\_HK}}\label{sec:Rp_HK}

The emission flux of \ion{Ca}{II} H and K lines is known as comprising the fluxes of stellar photosphere and chromosphere \citep{1984ApJ...276..254H, 1984ApJ...279..763N}. To acquire a purer chromospheric activity index, we should subtract the photospheric contribution from $R_{\rm HK}$. 
The photospheric and bolometric calibrated chromospheric activity index $R'_{\rm HK}$ \citep{1984ApJ...276..254H, 1984ApJ...279..763N} is defined as 
\begin{equation}
\label{eq:Rp_HK}
    R^\prime_{\rm HK} = R_{\rm HK} - R_{\rm phot},
\end{equation}
where $R_{\rm HK}$ has been described and derived in Section \ref{sec:RHK}, and $R_{\rm phot}$ represents the photospheric contribution which is the ratio between the photospheric flux and the bolometric flux
\begin{equation}\label{eq:Rphot_vs_Fphot}
    R_{\rm phot} = \frac{\mathcal{F}_{\rm phot}}{\sigma T^4_{\rm eff}}.
\end{equation}
$R_{\rm phot}$ can be derived from empirical spectral library (e.g., \citealt{1984ApJ...276..254H, 1984ApJ...279..763N, 2015MNRAS.452.2745S, 2018A&A...619A..73L}) or synthetic spectral library (e.g., \citealt{2013A&A...549A.117M, 2017A&A...600A..13A, 2023A&A...671A.162M}).
The value of ${\rm log} \,R_{\rm phot}$ in the literature that can be expressed by the polynomial form
\begin{equation}\label{eq:RHK_phot-5}
    {\rm log} \,R_{\rm phot} = \sum_{i=0}^{5} P_{i}X^{i}
\end{equation}
are presented in Table \ref{tab:Rphot}. In Equation \ref{eq:RHK_phot-5}, $X$ represents $B-V$ or $T_{\rm eff}$ and the $P_{i}$ (i=1,...,5) are the coefficients of the polynomial which are given in Table \ref{tab:Rphot}.

\citet{1984ApJ...279..763N} distilled the result of \citet{1984ApJ...276..254H} and expressed the relation between $\log\,R_{\rm phot}$ and $B-V$ via a cubic polynomial 
\begin{equation}\label{eq:Rphot-classic}
\log\,R_{\rm phot} = -4.898+1.918(B-V)^2-2.893(B-V)^3,
\end{equation}
for the main-sequence stars with $B-V > 0.44$ (see first row of Table \ref{tab:Rphot}).
They noted that $R_{\rm phot}$ becomes negligible for the case of $B-V \gtrsim 1.0$.
This expression of $R_{\rm phot}$ in Equation \ref{eq:Rphot-classic} was widely adopted to derive $R'_{\rm HK}$ in the majority of researches (e.g., \citealt{1996AJ....111..439H, 2004ApJS..152..261W, 2006AJ....132..161G, 2021A&A...646A..77G}), while a simpler linear form of $\log\,R_{\rm phot}$ is also available (used in \citealt{2007A&A...469..309C}).
By cross-match 72 stars with \citet{1984ApJ...279..763N}, \citet{2018A&A...619A..73L} fitted the $\log\,R_{\rm phot}$ into a formula of $T_{\rm eff}$
\begin{equation}\label{eq:Rphot_2018}
    \log \,R_{\rm phot} = -4.78845 - \frac{3.70700}{1 + (T_{\rm eff}/4598.92)^{17.527}},
\end{equation}
for stars in the range of $4350 \le T_{\rm eff} \le 6500\,{\rm K}$.
Based on the inactive stars observed by HARPS spectra, \citet{2015MNRAS.452.2745S} empirically fitted the $R_{\rm phot}$ for the main-sequence star in the range $0.4 \lesssim B-V \lesssim 1.9$, which expressed by an exponential function
\begin{equation}\label{eq:Rphot_2015}
    R_{\rm phot} = 1.48 \times 10^{-4} \cdot e^{-4.3658(B-V)}.
\end{equation}

\citet{2013A&A...549A.117M} adopted the synthetic spectra of PHOENIX to deduce the photospheric flux $\mathcal{F}_{\rm phot}$, which is expressed by a linear equation for main-sequence star in the range of $0.44 \le B-V < 1.28$ as
\begin{equation}\label{Fphot_2013}
\log\,\mathcal{F}_{\rm phot} = 7.49 - 2.06(B-V),
\end{equation}
which can be converted to $R_{\rm phot}$ by Equation \ref{eq:Rphot_vs_Fphot}.

 \citet{2017A&A...600A..13A} derived the $R_{\rm phot}$ in the range of $0.54 \le B-V \le 1.90$ based on the BT-Settl model \citep{2014IAUS..299..271A}.
Besides, \citet{2023A&A...671A.162M} employed the PHOENIX model to deduce a fifth-order equation that expresses the $\log\,R_{\rm phot}$ as a function of $T_{\rm eff}$.
For the same stellar $B-V$ or $T_{\rm eff}$ value, the values of $R_{\rm phot}$ deduced by synthetic spectra in \citet{2013A&A...549A.117M}, \citet{2017A&A...600A..13A} and \citet{2023A&A...671A.162M} are generally higher than the empirical calibration of \citet{1984ApJ...279..763N}.
\citet{2023A&A...671A.162M} thus introduced an offset 0.4612 to scale their result to \citet{1984ApJ...279..763N}.

\begin{table*} 
    \centering
    \caption{Coefficients for the $\log\,R_{\rm phot}$ expression in Equation \ref{eq:RHK_phot-5} in different researches. \label{tab:Rphot}}
    \begin{tabular}{cccccccc}

\hline \hline
Source & $P_0$ & $P_1$ & $P_2$ & $P_3$ & $P_4$ & $P_5$ & Valid Range \\
  &   &   &   &   &   &   &   of $B-V$ or $T_{\rm eff}$ \\
\hline
(1) & -4.898 & 0 & 1.918 & -2.893 & / & / & $0.44 < B-V$ \\
(2) & -3.749 & -1.036 & - 0.026 & -0.045 & / & / & $0.54 \le B-V \le 1.90$ \\
(3) & -37.550 & $0.032131$ & $-1.3177 \times 10^{-5}$ & $2.7133 \times 10^{-9}$ & $-2.7466 \times 10^{-13}$ & $1.0887 \times 10^{-17}$ & $2300 \le T_{\rm eff} \le 7200\,{\rm K}$ \\

\hline
    \end{tabular}
    \tablebib{
    (1)~\citet{1984ApJ...279..763N} (used for the calculation of $R'_{\rm HK,classic}$);
    (2)~\citet{2017A\string&A...600A..13A};
    (3)~\citet{2023A\string&A...671A.162M};
    }
\end{table*}

Same as $R_{\rm HK}$ described in Section \ref{sec:RHK}, we present two kinds of estimations of $R'_{\rm HK}$, denoted as $R'_{\rm HK,classic}$ and $R'_{\rm HK,PHOENIX}$, respectively. $R'_{\rm HK,classic}$ is calculated using $R_{\rm HK,classic}$ and the photospheric contribution derived from Equation \ref{eq:Rphot-classic} with $B-V$ estimated from Equation \ref{eq:teff_vs_bv}, while $R'_{\rm HK,PHOENIX}$ is computed based on $R_{\rm HK,PHOENIX}$ and the photospheric contribution $R_{\rm phot,PHOENIX}$ estimated as follows.

\begin{figure} 
    \resizebox{\hsize}{!}{\includegraphics{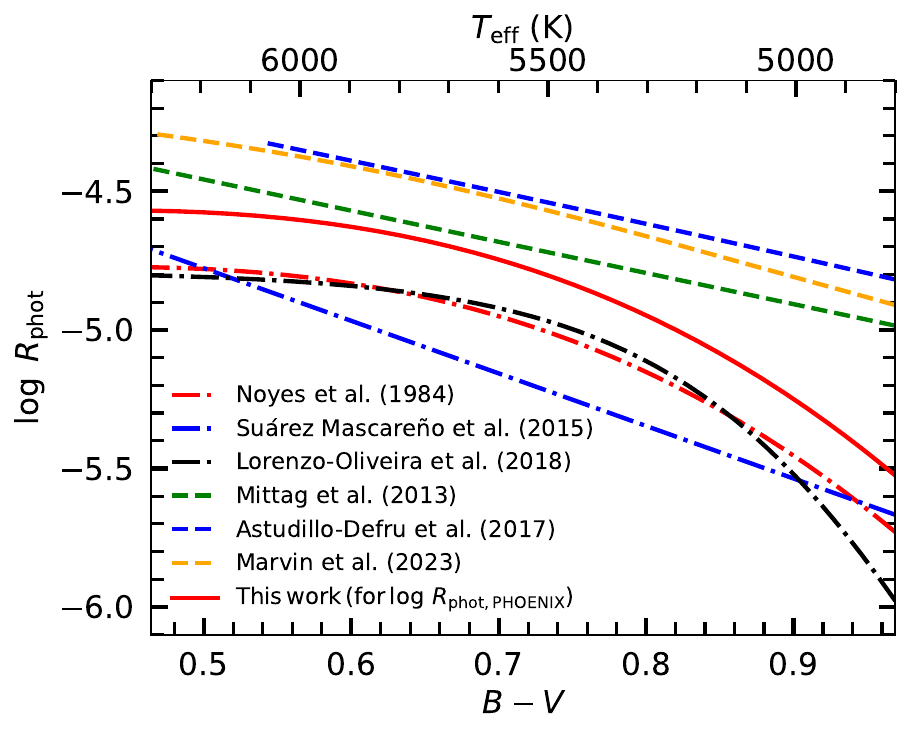}}
    \caption{Relations between $\log\,R_{\rm phot}$ and $B-V$ (or $T_{\rm eff}$) adopted in different researches. The formulas derived from empirical and synthetic spectral libraries in the literature are indicated by dashed-dotted and dashed lines respectively. 
    The formula of $\log\,R_{\rm phot,PHOENIX}$ adopted in this work is indicated by the red solid line.
    \label{fig:Rphot-teff}}
\end{figure}

Because the values of $R_{\rm HK,PHOENIX}$ are approximately $\beta$ times larger than the values of $R_{\rm HK,classic}$, we introduce a $\beta$-coefficient to scale $R_{\rm phot,classic}$ to $R_{\rm phot,PHOENIX}$ and the corresponding $\log\,R_{\rm phot,PHOENIX}$ can be expressed by
\begin{equation}\label{eq:Rphot-PHOENIX}
\begin{aligned}
    \log\,R_{\rm phot,PHOENIX} &= \log(\beta \cdot R_{\rm phot,classic}) \\
    &=-4.694+1.918(B-V)^2-2.893(B-V)^3,
\end{aligned}
\end{equation}
for $B-V>0.44$.
In Figure \ref{fig:Rphot-teff}, we present the relations between $\log\,R_{\rm phot}$ and $B-V$ (or $T_{\rm eff}$) adopted in different researches. 
As discussed above, the $R_{\rm phot,PHOENIX}$ is scaled from the results of \citet{1984ApJ...279..763N} using a scale factor $\beta$ related to the method based on the PHOENIX model. Hence, the red solid curve in Figure \ref{fig:Rphot-teff} differs from those obtained in \citet{2013A&A...549A.117M} and \citet{2023A&A...671A.162M}.

Since the detailed stellar atmospheric parameters ($T_{\rm eff}$, $\log\,g$ and [Fe/H]) are available for LAMOST, we estimate the $B-V$ in Equation \ref{eq:Rphot-PHOENIX} by considering not only $T_{\rm eff}$, but also the stellar atmospheric parameters $\log\,g$ and [Fe/H].
Based on the InfraRed Flux Method, \citet{2010A&A...512A..54C} found that there is very little dependence of $B-V$ on $\log\,g$ and provided a relation between $T_{\rm eff}$, $B-V$,  and [Fe/H], with $B-V$ and [Fe/H] in the range of $0.18 \le B-V \le 1.29$ and $-5.0 \le {\rm [Fe/H]} \le 0.4\,{\rm dex}$.
We examine the extendability of the [Fe/H] upper boundary and still employ the relation to obtain the $B-V$ for the small amount of spectra with [Fe/H] slightly exceeding $0.4\,{\rm dex}$.
After obtaining $B-V$ from $T_{\rm eff}$ and [Fe/H], we then estimate the photospheric contribution $R_{\rm phot,PHOENIX}$ based on Equation \ref{eq:Rphot-PHOENIX}. 
Because both the $R_{\rm HK,PHOENIX}$ and $R_{\rm phot,PHOENIX}$ are about $\beta$ times higher than the corresponding classic indexes, to be consistent with classic studies, we calculated the $R'_{\rm HK,PHOENIX}$ by 
\begin{equation}
\label{eq:Rp_HK_PHOENIX}
    R^\prime_{\rm HK,PHOENIX} = \frac{1}{\beta}(R_{\rm HK,PHOENIX} - R_{\rm phot,PHOENIX}).
\end{equation}

\begin{figure} 
    \resizebox{\hsize}{!}{\includegraphics{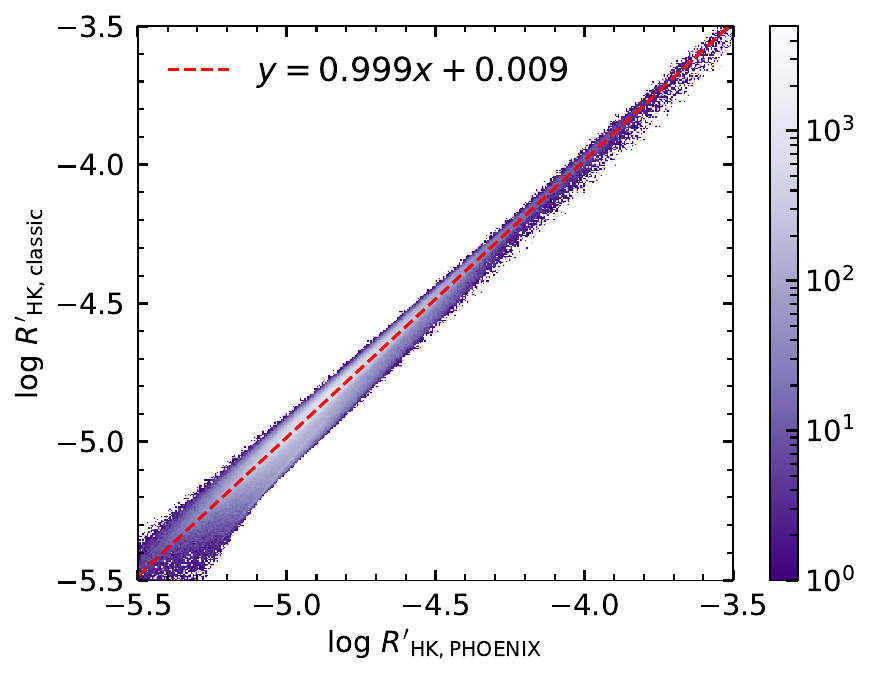}}
    \caption{Scatter diagram between $\log\,R'_{\rm HK,PHOENIX}$ and $\log\,R'_{\rm HK,classic}$ for the LAMOST LRS spectra of solar-like stars studied in this work. The color scale represents the density of data points.
    The dashed line is the linear fit between $\log\,R'_{\rm HK,PHOENIX}$ and $\log\,R'_{\rm HK,classic}$.}
    \label{fig:Rp_HK_PHOENIX-Rp_HK_classic}
\end{figure}

In Figure \ref{fig:Rp_HK_PHOENIX-Rp_HK_classic}, we present a comparison between the values of the two indexes $\log\,R'_{\rm HK,PHOENIX}$ and $\log\,R'_{\rm HK,classic}$ for the LAMOST LRS spectra of solar-like stars studied in this work.
As shown in Figure \ref{fig:Rp_HK_PHOENIX-Rp_HK_classic}, there exists a linear correlation between these two quantities, the fitting formula is
\begin{equation}\label{eq:Rp_HK_PHOENIX_vs_Rp_HK_classic}
    \log\,R'_{\rm HK,classic} = 0.999\log\,R'_{\rm HK,PHOENIX}+0.009.
\end{equation}

\citet{2017ApJ...835...25E} estimated the average value of $\log\,R'_{\rm HK,\odot}$ as $-4.9427 \pm 0.0072$ based on the $S_{\rm MWO,\odot}$ in the 15$-$24 solar cycle and $(B-V)_{\odot} = 0.653 \pm 0.003$.
Taking $T_{\rm eff} = 5777\,{\rm K}$, $\log\,g=4.44\,{\rm dex}$, ${\rm [Fe/H]} = 0.0\,{\rm dex}$ and the same solar $B-V$, we can obtain the $\log\,R'_{\rm HK,PHOENIX}=-4.9599 \pm 0.0051$ for the Sun.

\subsection{Estimation of the Uncertainty of Chromospheric Activity Indexes}\label{sec:estiamtion_uncertainty}

We estimated the uncertainties of $\log\,R_{\rm HK,classic}$, $\log\,R'_{\rm HK,classic}$, $\log\,R_{\rm HK,PHOENIX}$, and $\log\,R'_{\rm HK,PHOENIX}$  with the Monte Carlo error propagation. Because the $\log\,R_{\rm HK,PHOENIX}$ values are calculated by Equation \ref{eq:RHK_vs_F_RV}, the uncertainties of $\log\,R_{\rm HK,PHOENIX}$ are yielded from the uncertainties of $S_{\rm MWO}$ and $\mathcal{F}_{\rm RV}$.
The uncertainties of $\log\,R_{\rm HK,classic}$ predominantly arise from the uncertainties of $S_{\rm MWO}$ and $C_{\rm cf}$ as presented in Equation \ref{eq:RHK_vs_Ccf}.
As illustrated in Paper I, we estimated the uncertainties of $S_{\rm MWO}$ by considering the impact of the uncertainties of the spectral flux, the discretization in spectral data, and the uncertainty of radial velocity.
Regarding the uncertainties of $\mathcal{F}_{\rm RV}$, it is affected by the uncertainties of stellar atmospheric parameters $T_{\rm eff}$, $\log\,g$ and [Fe/H] due to Equation \ref{eq:F_RV_vs_teff-logg-feh}.
Since we calculate the value of $C_{\rm cf}$ through the $B-V$ value derived from Equation \ref{eq:teff_vs_bv}, the uncertainties of $C_{\rm cf}$ are influenced by the uncertainties of $B-V$ which is propagated from the uncertainties of $T_{\rm eff}$.

\begin{figure*} 
    \resizebox{\hsize}{!}{\includegraphics{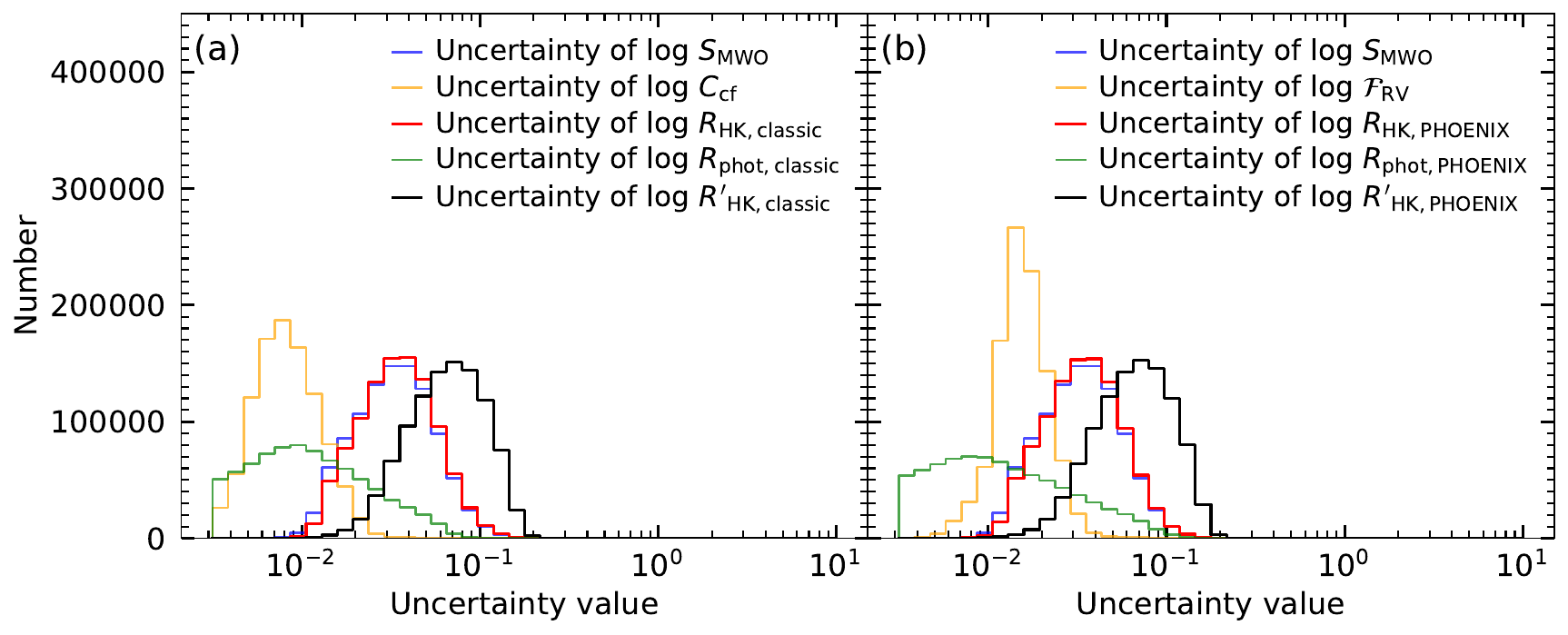}}
    \caption{Histograms of the uncertainties of the parameters derived from (a) the method in the classic literature and (b) the method based on the PHOENIX model.}
    \label{fig:Rindex_err_hist}
\end{figure*}

Figure \ref{fig:Rindex_err_hist}(a) illustrates the histograms of the uncertainties of $\log\,S_{\rm MWO}$, $\log\,C_{\rm cf}$, $\log\,R_{\rm HK,classic}$, $\log\,R_{\rm phot,classic}$ and $\log\,R'_{\rm HK,classic}$, while Figure \ref{fig:Rindex_err_hist}(b) shows the uncertainties for $\log\,S_{\rm MWO}$, $\log\,\mathcal{F}_{\rm RV}$, $\log\,R_{\rm HK,PHOENIX}$, $\log\,R_{\rm phot,PHOENIX}$ and $\log\,R'_{\rm HK,PHOENIX}$.
As shown in Figure \ref{fig:Rindex_err_hist}, the uncertainties of $\log\,R_{\rm HK,PHOENIX}$ and $\log\,R_{\rm HK,classic}$ are both primarily governed by the uncertainties of $S_{\rm MWO}$.
The uncertainties of $\log\,S_{\rm MWO}$, $\log\,R_{\rm HK,classic}$, $\log\,R'_{\rm HK,classic}$, $\log\,R_{\rm HK,PHOENIX}$ and $\log\,R'_{\rm HK,PHOENIX}$ are distributed around 0.030, 0.030, 0.065, 0.030 and 0.065 respectively.

\section{Results and Discussion} \label{sec:results}

\subsection{Stellar Chromospheric Activity Database}

In Section \ref{sec:process}, we investigate the stellar chromospheric activity through two kinds of methods. The chromospheric activity parameters derived from the method in the classic literature are denoted with {\tt\string classic}, while those derived from the method based on the PHOENIX model are denoted with {\tt\string PHOENIX}.
We provide the database of chromospheric activity parameters for 1,122,495 LAMOST LRS spectra of solar-like stars, which is available at \url{https://doi.org/10.5281/zenodo.8378849} (compiled in a CSV format file: {\tt\string CaIIHK\_Activity\_Indexes\_LAMOST\_DR8\_LRS.csv}).
The database mainly includes the chromospheric activity parameters $S_{\rm tri}$, $S_{\rm MWO}$, $R_{\rm HK,classic}$, $R'_{\rm HK,classic}$, $R_{\rm HK,PHOENIX}$ and $R'_{\rm HK,PHOENIX}$, as well as their uncertainties.
The columns in the catalog of the database are presented in Table \ref{tab:catalog-columns}.

\begin{table*} 
	\centering
	\caption{Columns in the catalog of the database. \label{tab:catalog-columns}}
	\begin{tabular}{llccll}
		\hline\hline
		\multicolumn{1}{l}{}
		& \multicolumn{1}{l}{Column}
		& \multicolumn{1}{l}{}
		& \multicolumn{1}{l}{Unit}
		& \multicolumn{1}{l}{}
		& \multicolumn{1}{l}{Description} \\
		\hline
		& {\tt\string obsid} &&  && Unique observation identifier of LAMOST LRS spectrum \\
		& {\tt\string obsdate} &&  && Spectral observation date \\
		& {\tt\string fitsname} && && FITS file name of LAMOST LRS spectrum\\
		& {\tt\string snrg} && && Signal-to-noise ratio in the $g$ band ($\mathrm{S/N}_g$) of LAMOST LRS spectrum\\
		& {\tt\string snrr} && && Signal-to-noise ratio in the $r$ band ($\mathrm{S/N}_r$) of LAMOST LRS spectrum\\
		& {\tt\string teff} && K && Effective temperature ($T_\mathrm{eff}$) derived from the LASP\\
		& {\tt\string teff\_err} && K && Uncertainty of $T_\mathrm{eff}$ derived from the LASP\\
		& {\tt\string logg} && dex && Surface gravity ($\log\,g$) derived from the LASP\\
		& {\tt\string logg\_err} && dex && Uncertainty of $\log\,g$ derived from the LASP\\
		& {\tt\string feh} && dex && Metallicity ([Fe/H]) derived from the LASP\\
		& {\tt\string feh\_err} && dex && Uncertainty of [Fe/H] derived from the LASP\\
		& {\tt\string rv} && km/s && Radial velocity ($V_r$) derived from the LASP\\
		& {\tt\string rv\_err} && km/s && Uncertainty of $V_r$ derived from the LASP\\
		& {\tt\string ra\_obs} && degree && Right ascension (RA) of fiber pointing \\
		& {\tt\string dec\_obs} && degree && Declination (DEC) of fiber pointing \\
		& {\tt\string gaia\_source\_id} && && Source identifier in Gaia DR3 catalog\\
		& {\tt\string gaia\_g\_mean\_mag} && && $G$ magnitude in Gaia DR3 catalog\\
		& {\tt\string S\_tri}* && && $S_{\rm tri}$ index  of LAMOST LRS \\ 
		& {\tt\string S\_tri\_err}* && && Uncertainty of $S_{\rm tri}$ \\ 
		& {\tt\string S\_MWO}* && && $S_{\rm MWO}$ \\ 
		& {\tt\string S\_MWO\_err}* && && Uncertainty of $S_{\rm MWO}$ \\ 
		& {\tt\string log\_R\_HK\_classic}* && && $\log\,R_{\rm HK,classic}$ \\
		& {\tt\string log\_R\_HK\_classic\_err}* && && Uncertainty of $\log\,R_{\rm HK,classic}$ \\
		& {\tt\string log\_R\_p\_HK\_classic}* && && $\log\,R'_{\rm HK,classic}$ \\
		& {\tt\string log\_R\_p\_HK\_classic\_err}* && && Uncertainty of $\log\,R'_{\rm HK,classic}$ \\
		& {\tt\string log\_R\_HK\_PHOENIX}* && && $\log\,R_{\rm HK,PHOENIX}$ \\
		& {\tt\string log\_R\_HK\_PHOENIX\_err}* && && Uncertainty of $\log\,R_{\rm HK,PHOENIX}$ \\
		& {\tt\string log\_R\_p\_HK\_PHOENIX}* && && $\log\,R'_{\rm HK,PHOENIX}$ \\
		& {\tt\string log\_R\_p\_HK\_PHOENIX\_err}* && && Uncertainty of $\log\,R'_{\rm HK,PHOENIX}$ \\
		\hline
	\end{tabular}
	\tablefoot{The columns denoted by an asterisk represent the parameters estimated in this work, and the remaining columns utilized in our research are collected from the data release of LAMOST DR8 V2.0. The detailed definitions of the chromospheric activity parameters can be seen in Section \ref{sec:process}.}
\end{table*}

The $\log\,R'_{\rm HK,classic}$ and $\log\,R'_{\rm HK,PHOENIX}$ values of 743 and 821 spectra, respectively, are not available (recorded as '{\tt\string -9999}' in the database).
One of the reason is the value of stellar parameters exceeds the valid range of the empirical formula of $R_{\rm phot}$ (0 and 13 spectra for {\tt\string classic} and {\tt\string PHOENIX}, respectively).
The other reason is that the estimated value of photospheric contribution is large than the value of $R_{\rm HK}$ for very few spectra (743 and 808 spectra for {\tt\string classic} and {\tt\string PHOENIX}, respectively).
This situation would occur because the photospheric contributions are determined empirically, leading to overestimations for some stars; or there are uncertainties in the evaluation of $R_{\rm HK}$.
These spectra are not involved in the subsequent discussion.
In Sections \ref{sec:RHK} and \ref{sec:Rp_HK}, we have performed a comparison between $\log\,R_{\rm HK,classic}$ and $\log\,R_{\rm HK,PHOENIX}$, and between $\log\,R'_{\rm HK,classic}$ and $\log\,R'_{\rm HK,PHOENIX}$. The results indicate that $\log\,R_{\rm HK,PHOENIX}$ and $\log\,R'_{\rm HK,PHOENIX}$ are approximately linearly correlated with $\log\,R_{\rm HK,classic}$ and $\log\,R'_{\rm HK,classic}$, respectively (see Figures \ref{fig:R_HK_PHOENIX-R_HK_classic} and \ref{fig:Rp_HK_PHOENIX-Rp_HK_classic}).
In the next subsection, we discuss the distribution of chromospheric activity primarily based on $R_{\rm HK,PHOENIX}$ and $R'_{\rm HK,PHOENIX}$.

\subsection{Distribution of Chromospheric Activity Index}

Among the 1,122,495 LAMOST LRS spectra of solar-like stars, there are 861,505 stars with 'gaia\_source\_id' available in 
{\tt\string LAMOST LRS AFGK Catalog}. In this section, we investigate the distribution of chromospheric activity index based on these stars. If a star is recorded more than once in our dataset, we use the median values of the chromospheric activity parameters from the multiple observed spectra.
In Paper I, we have calibrated the $S$ index of LAMOST to $S_{\rm MWO}$, and we also compare the $R'_{\rm HK}$ with the results in the literatures, as illustrated in Appendix \ref{sec:calibration}.
There is an approximate consistency between our $R'_{\rm HK}$ values and those from other instruments for the common targets.

\begin{figure} 
    \resizebox{\hsize}{!}{\includegraphics{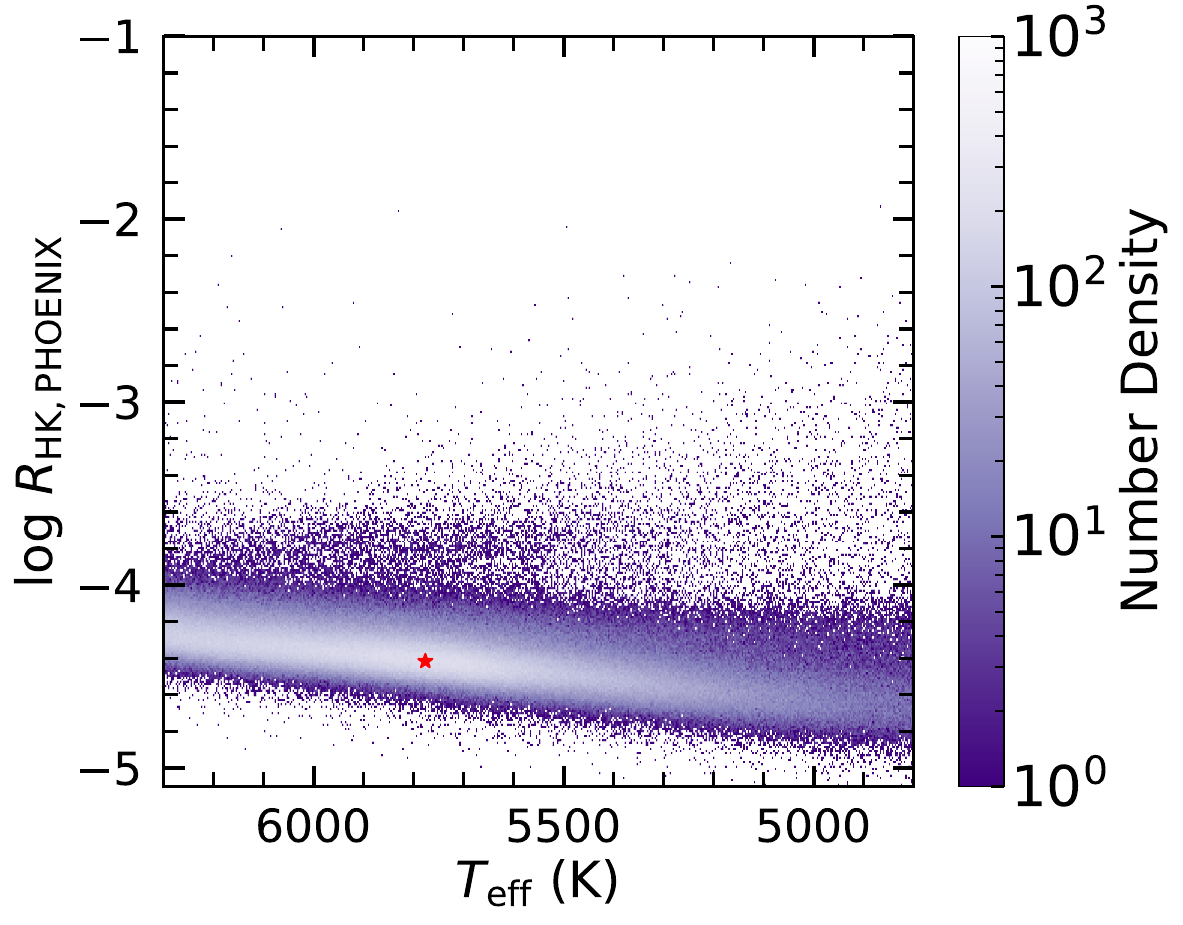}}
    \caption{Scatter diagram of $\log\,R_{\rm HK,PHOENIX}$ with $T_{\rm eff}$ for all  of solar-like stars investigated in this work. The color scale represents the density of data points. The $R_{\rm HK,PHOENIX}$ value of the Sun is marked with a ($\star$) symbol.
    }
    \label{fig:R_HK_PHOENIX-teff}
\end{figure}

In Figure \ref{fig:R_HK_PHOENIX-teff}, we display the distribution of $\log\,R_{\rm HK,PHOENIX}$ with $T_{\rm eff}$ for the 861,505 solar-like stars.
The solar value of $\log\,R_{\rm HK,PHOENIX}$ ($-4.416 \pm 0.001$) is displayed in Figure \ref{fig:R_HK_PHOENIX-teff} with a '$\star$' symbol, which is calculated by Equation \ref{eq:RHK_vs_FHK} with the solar $\mathcal{F}_{\rm HK,PHOENIX} = (2.423 \pm 0.007) \times 10^{6}$ ${\rm erg}\,{\rm cm}^{-2}\,{\rm s}^{-1}$ given in Section \ref{sec:RHK}.

\begin{figure*} 
    \resizebox{\hsize}{!}{\includegraphics{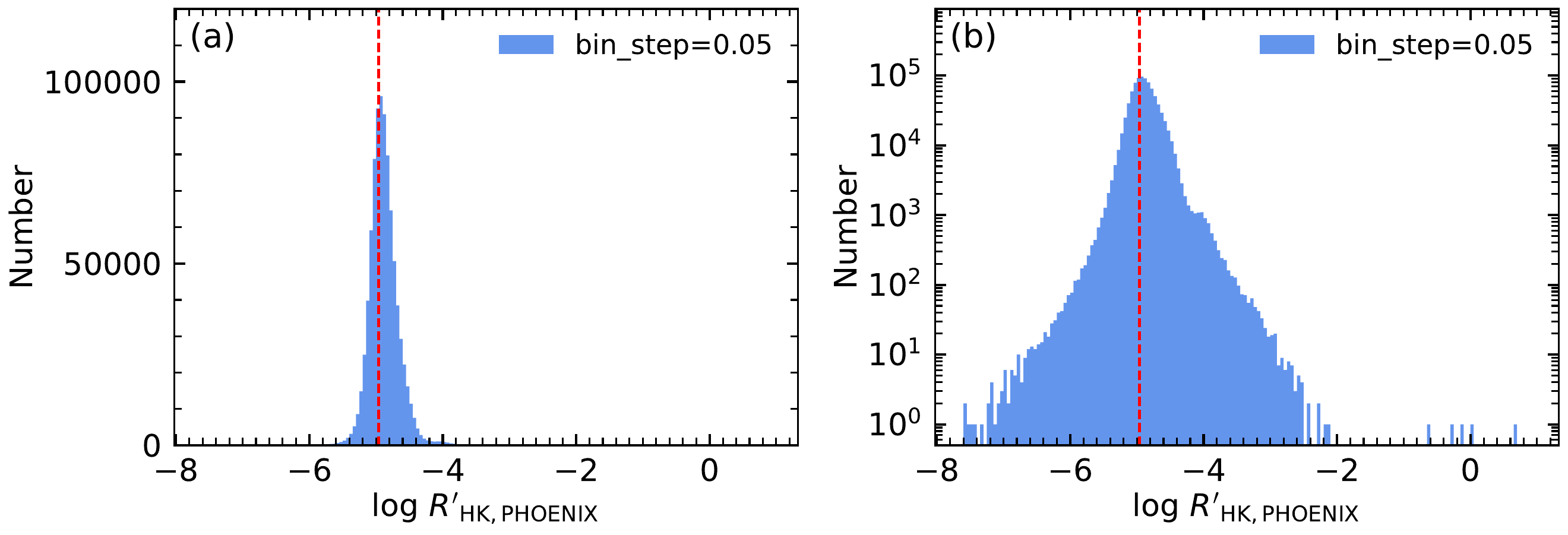}}
    \caption{Histograms of the $\log\,R'_{\rm HK,PHOENIX}$ values for the  solar-like stars investigated in this work through (a) linear-scale and (b) logarithmic-scale vertical axes. 
    The $\log\,R'_{\rm HK,PHOENIX}$ value of the Sun (-4.9599 as given in Section \ref{sec:Rp_HK}) is indicated by a vertical dashed line.}
    \label{fig:Rp_HK_PHOENIX_hist}
\end{figure*}

It is not surprising that the distribution trend of $\log\,R_{\rm HK,PHOENIX}$ shows a clear correlation with $T_{\rm eff}$.
Although the $R_{\rm HK,PHOENIX}$ is the bolometric calibration of the surface flux, the photospheric contribution related to stellar spectral types is still contained.
As mentioned in Section \ref{sec:Rp_HK}, it is necessary to further remove the contribution of photosphere to obtain $R'_{\rm HK,PHOENIX}$.
The histograms of the $\log\,R'_{\rm HK,PHOENIX}$ values are exhibited in Figures \ref{fig:Rp_HK_PHOENIX_hist}(a) and (b) in linear-scale and logarithmic-scale vertical axises, respectively. 
The peak of the distribution is at about -4.90.
This peak value is close to the solar $\log\,R'_{\rm HK,PHOENIX}=-4.9599$ as given in Section \ref{sec:Rp_HK}.

The Vaughan-Preston gap (VP gap, \citealt{1980PASP...92..385V}), known as the bimodal distribution of chromospheric activity, is not observed in Figure \ref{fig:Rp_HK_PHOENIX_hist}.
The separation of $\log\,R'_{\rm HK}$ values between active and inactive stars may suggest the existence of different dynamo mechanisms \citep{1980PASP...92..385V, 1984ApJ...279..763N, 1996AJ....111..439H, 2006MNRAS.372..163J, 2006AJ....132..161G, 2021A&A...646A..77G}.
\citet{2018A&A...616A.108B} investigated a global sample of 4451 cool stars from high-resolution HARPS spectra and concluded that the VP gap is not pronounced. 
A significant proportion of the stars have intermediate activity levels around $\log\,R'_{\rm HK}=-4.75$ in \citet{2018A&A...616A.108B}.
In contrast, the bimodal distribution of chromospheric activity in \citet{2021A&A...646A..77G} is relatively significant, based on 1674 F-, G- and K-type stars from the HARPS sample.
\citet{2006AJ....132..161G} and \citet{2017ApJ...848...34H} proposed that the VP gap tends to appear for stars with [Fe/H] greater than -0.2, which is inflexible for stars in this work.
The VP gap is also influenced by the rotation rate (e.g. \citealt{1984ApJ...279..763N, 1987A&A...177..131R}), and the relationship between rotation and stellar chromospheric activity  in LAMOST samples will be investigated in the future.
\citet{1980PASP...92..385V} suggested that the VP gap might originate from different dynamo mechanisms or statistical bias.
The absence of VP gap in the distribution of chromospheric activity for our solar-like stars could be attributed to three possible factors: 1) a gradual diminishing of chromospheric activity during the evolution of solar-like stars; 2) the influence of different stellar properties on the bimodal distribution of the chromospheric activity within our samples, which should be explored in more detail in the future, or 3) the loss of some information in the spectral profile due to the limited resolution of LAMOST LRS spectra \citep{2011A&A...531A...8J}.

\begin{figure*} 
    \resizebox{\hsize}{!}{\includegraphics{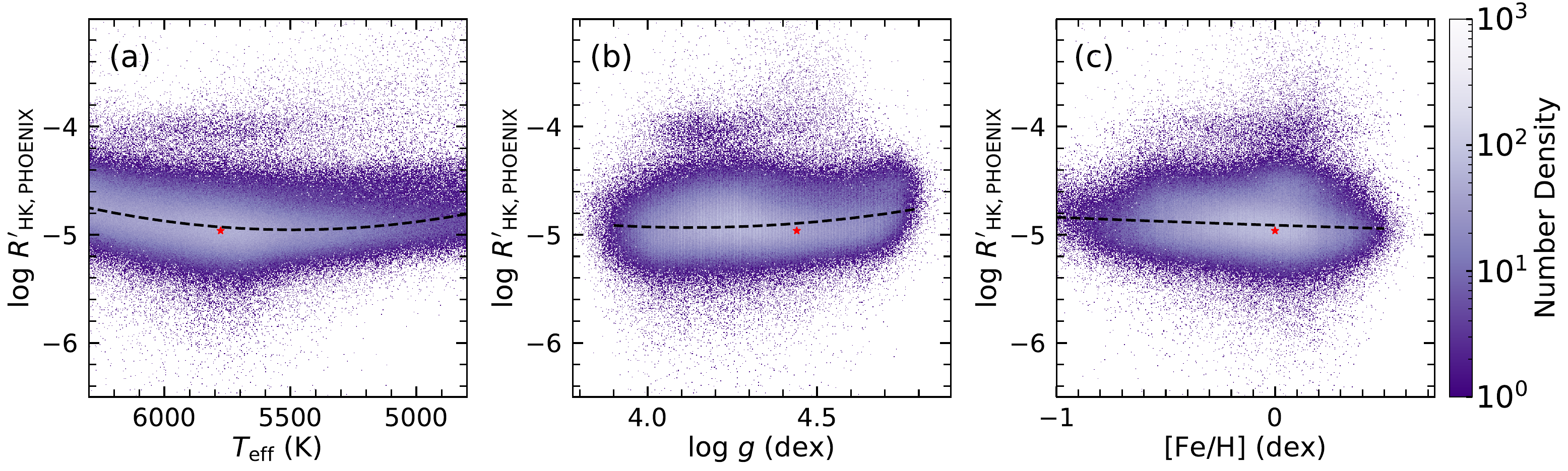}}
    \caption{Scatter diagrams of $\log\,R'_{\rm HK,PHOENIX}$ with (a) $T_{\rm eff}$, (b) $\log\,g$, and (c) [Fe/H] for the solar-like stars investigated in this work. The position of the solar $\log\,R'_{\rm HK,PHOENIX}$ in these diagrams is denoted by a ($\star$) symbol.
    The black dashed lines represent the fitted median values of $\log\,R'_{\rm HK,PHOENIX}$. The color scale indicates the number density.
    Some values of $\log\,R'_{\rm HK,PHOENIX}$ outside the displayed range are considered insignificant for the overall distribution and are therefore not shown.
	}
    \label{fig:Rp_HK_PHOENIX-teff-logg-feh}
\end{figure*}

We display the distributions of $\log\,R'_{\rm HK,PHOENIX}$ with $T_{\rm eff}$, $\log\,g$ and [Fe/H] in Figures \ref{fig:Rp_HK_PHOENIX-teff-logg-feh}(a), (b) and (c), respectively.
 To show the trends of $\log\,R'_{\rm HK,PHOENIX}$ with these stellar atmospheric parameters, the $\log\,R'_{\rm HK,PHOENIX}$ values are homogeneously segregated into equal-width bins for $4800 < T_{\rm eff} < 6300\,{\rm K}$, $3.9<\log\,g<4.8\,{\rm dex}$ and $\rm -1.0<[Fe/H]<0.6\, dex$ with steps of $50\,{\rm K}$, $0.1 \,{\rm dex}$ and $0.1\,{\rm dex}$, respectively;
and the fitted median values of the $\log\,R'_{\rm HK,PHOENIX}$ in each bin with $T_{\rm eff}$, $\log\,g$ and [Fe/H] are marked by the black dashed lines in Figure \ref{fig:Rp_HK_PHOENIX-teff-logg-feh}.
The formulas of these fitted trends are expressed by the following quadratic polynomials
\begin{equation}\label{eq:log_Rp_HK_PHOENIX-teff}
    \log\,R'_{\rm HK,PHOENIX} = 4.367
    - 3.397 \times 10^{-3} T_{\rm eff}
    +3.094 \times 10^{-7} T^2_{\rm eff} 
    ,
\end{equation}
\begin{equation}\label{eq:log_Rp_HK_PHOENIX-logg}
    \log\,R'_{\rm HK,PHOENIX} = 1.595
    -3.166 \log\,g
    +0.384 (\log\,g)^2  ,
\end{equation}
\begin{equation}\label{eq:log_Rp_HK_PHOENIX-feh}
    \log\,R'_{\rm HK,PHOENIX} = -4.912
    -0.065 {\rm [Fe/H]} 
    +7.894 \times 10^{-3} {\rm [Fe/H]}^2 .
\end{equation}
As shown in Figure \ref{fig:Rp_HK_PHOENIX-teff-logg-feh}, the median values of $\log\,R'_{\rm HK,PHOENIX}$ with $T_{\rm eff}$  have a minimum at about $T_{\rm eff}=5500\,{\rm K}$, while the dependence of the median values of $\log\,R'_{\rm HK,PHOENIX}$ on  $\log\,g$ and [Fe/H] is relatively weak.
Besides, it can be seen that the solar $\log\,R'_{\rm HK,PHOENIX}$ value is approximately on the fitting lines of the median $\log\,R'_{\rm HK,PHOENIX}$ values for all the three parameters, and the value of solar chromospheric activity index is located at the midpoint of the solar-like star sample. This result based on our extensive archive support the view that the dynamo mechanism of solar-like stars is generally consistent with the Sun.

\citet{1996AJ....111..439H} classified the stellar chromospheric activity into four levels: very active ($\log\,R'_{\rm HK}$ larger than -4.20), active ($\log\,R'_{\rm HK}$ from -4.75 to -4.20), inactive ($\log\,R'_{\rm HK}$ from -5.10 to -4.75) and very inactive ($\log\,R'_{\rm HK}$ less than -5.10).
Following this classification, based on the $\log\,R'_{\rm HK,PHOENIX}$ values of 861,505 stars, we can obtain the proportions of very active, active, inactive and very inactive solar-like stars as 1.03\%, 21.68\%, 65.27\% and 12.03\%, respectively. While for the values of $\log\,R'_{\rm HK,classic}$, the proportions are 1.07\%, 24.53\%, 62.98\% and 11.41\%, respectively.
The proportions of stars in the different stellar chromospheric activity classes are 2.6\%, 27.1\%, 62.5\% and 7.9\% in \citet{1996AJ....111..439H}, and 1.2\%, 28.5\%, 66.9\% and 3.5\% in \citet{2021A&A...646A..77G}.
When using a threshold of $\log\,R'_{\rm HK}=-4.75$ to classify stars as active and inactive, \citet{1996AJ....111..439H} and \citet{2021A&A...646A..77G} found that 29.7\% stars are classified as active.
Classifying the solar-likes stars studied in this work with $\log\,R'_{\rm HK,PHOENIX}=-4.75$ and $\log\,R'_{\rm HK,classic}=-4.75$, we can obtain the proportions of active solar-like stars as 22.71\% and 25.60\%, respectively.
The proportions are relatively consistent with the results of \citet{1996AJ....111..439H} and \citet{2021A&A...646A..77G}.

\begin{figure*} 
    \centering
    \resizebox{\hsize}{!}{\includegraphics{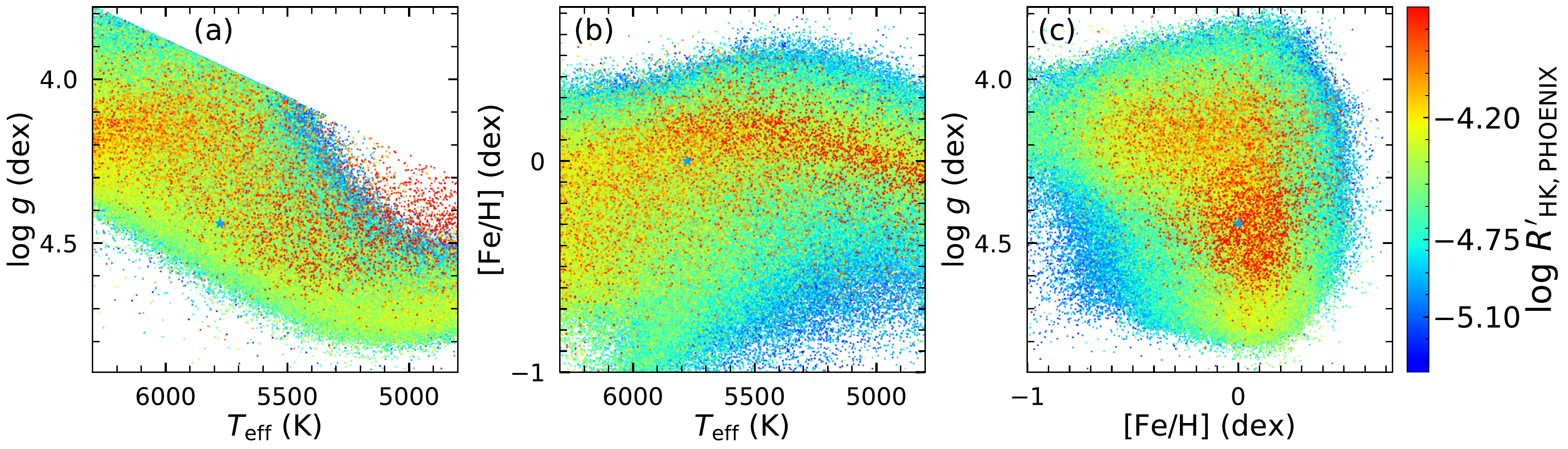}}
    \caption{Distributions of $\log\,R'_{\rm HK,PHOENIX}$ values in (a) $T_{\rm eff}$ vs. $\log\,g$, (b) $T_{\rm eff}$ vs. [Fe/H], and (c) [Fe/H] vs. $\log\,g$ parameter spaces for the solar-like stars investigated in this work.
    The stellar chromospheric activity levels of very active, active, inactive and very inactive are indicated by different colors. The values of $\log\,R'_{\rm HK,PHOENIX}$ that fall outside the range of the color bar are represented by the boundary value.
    The data points with smaller $\log\,R'_{\rm HK,PHOENIX}$ are overlaid by the data points with larger $\log\,R'_{\rm HK,PHOENIX}$. 
    The positions of the solar $\log\,R'_{\rm HK,PHOENIX}$ are marked in the diagrams with a ($\star$) symbol colored in accordance with the solar activity level.
    }
    \label{fig:Rp_HK_PHOENIX-teff-logg-feh_color}
\end{figure*}

\begin{figure} 
    \resizebox{\hsize}{!}{\includegraphics{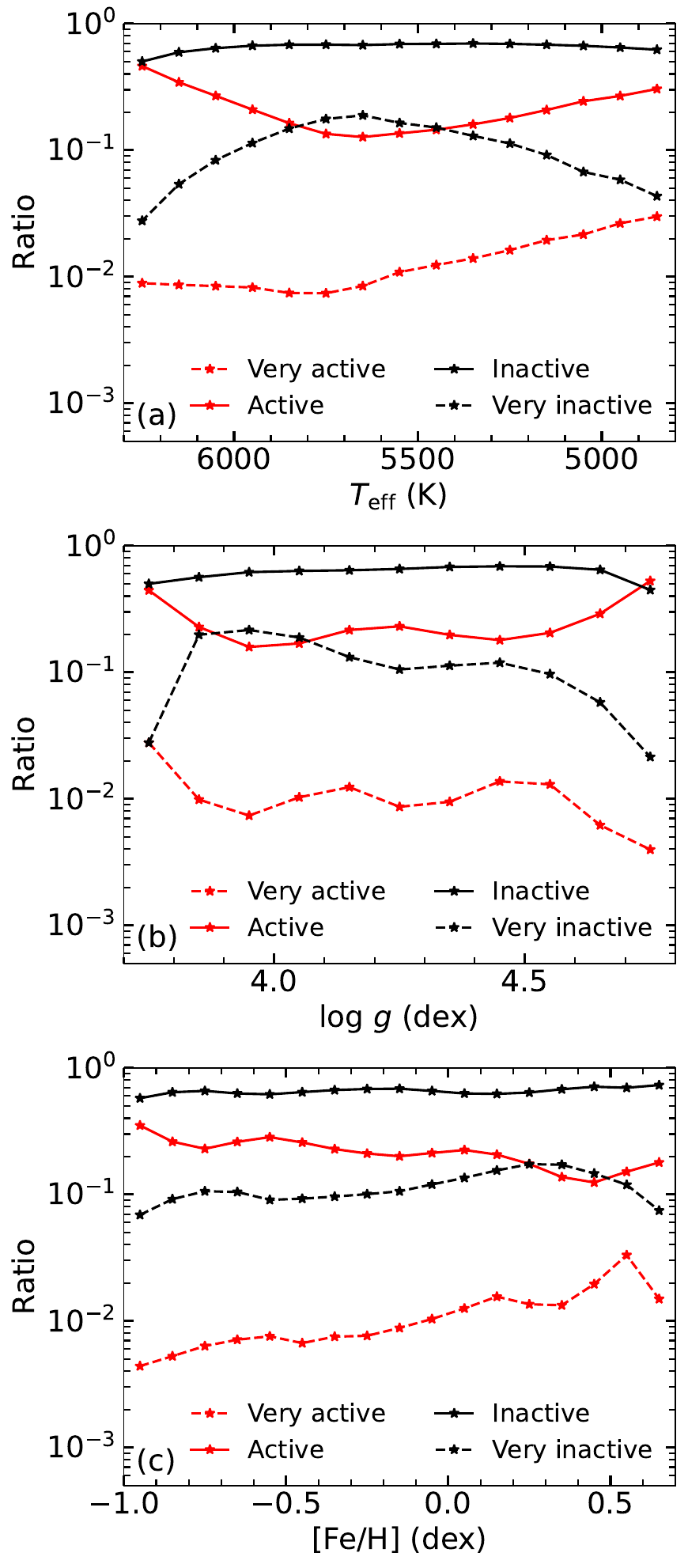}}
    \caption{Relations of the proportions of very active (red dashed line), active (red line), inactive (black line) and very inactive (black dashed line) solar-like stars with (a) $T_{\rm eff}$, (b) $\log\,g$ and (c) [Fe/H]. The proportions values are obtained by dividing the $T_{\rm eff}$, $\log\,g$ and [Fe/H] into bins with step size of $100\,{\rm K}$, $0.1\,{\rm dex}$ and $0.1\,{\rm dex}$, respectively, and the central values of each bin are used to represent the corresponding stellar atmospheric parameters.}
    \label{fig:ratio}
\end{figure}

In Figure \ref{fig:Rp_HK_PHOENIX-teff-logg-feh_color}, we show the distributions of $\log\,R'_{\rm HK,PHOENIX}$ values in the $T_{\rm eff}$ vs. $\log\,g$, $T_{\rm eff}$ vs. [Fe/H], and  [Fe/H] vs. $\log\,g$ parameters spaces.
The stellar chromospheric activity levels of very active, active, inactive, and very inactive are indicated by different colors.
It can be seen from Figure \ref{fig:Rp_HK_PHOENIX-teff-logg-feh_color} that the higher the stellar chromospheric activity levels, the narrower the distribution areas in the parameters spaces.
Since the LAMOST LRS spectra of solar-like stars with determined stellar atmospheric parameters are sufficient, we further investigate the relations between the proportions of solar-like stars with different chromospheric activity levels (classified by $R'_{\rm HK,PHOENIX}$) and the stellar atmospheric parameters ($T_{\rm eff}$, $\log\,g$ and [Fe/H]). The proportions of very active, active, inactive and very inactive solar-like stars with different stellar atmospheric parameters are shown in Figure \ref{fig:ratio}.
The proportions values in Figure \ref{fig:ratio} are obtained by dividing the $T_{\rm eff}$, $\log\,g$ and [Fe/H] into bins with step size of $100\,{\rm K}$, $0.1\,{\rm dex}$ and $0.1\,{\rm dex}$, respectively; and the central values of each bin are used to represent the corresponding stellar atmospheric parameters.

Figure \ref{fig:ratio}(a) shows that the proportions of inactive solar-like stars exhibit a relatively stable trend within the $T_{\rm eff}$ range of 4800 to $6000 \,{\rm K}$.
For the very inactive solar-like stars, there is an increasing trend in the proportions as the $T_{\rm eff}$ decreases within the $T_{\rm eff}$ range from 6300 to $5650\, {\rm K}$, and the proportions decrease within the $T_{\rm eff}$ range from 5650 to $4800\, {\rm K}$. 
In contrast, the proportions of active solar-like stars exhibit a decreasing trend with decreasing $T_{\rm eff}$ from 6300 to $5650\, {\rm K}$, and the decreasing trend of the proportions of active solar-like stars is reversed for $T_{\rm eff}$ lower than $5650\, {\rm K}$.
The proportions of very active solar-like stars are almost stable for $T_{\rm eff}>5900\,{\rm K}$, while they increase for $T_{\rm eff}<5600\,{\rm K}$. The minimum value of the proportions of very active solar-like stars is around $T_{\rm eff}=5700\,{\rm K}$.
Based on the proportions of active and very active solar-like stars, we conclude that the occurrence rate of high levels of chromospheric activity is lower among the stars with effective temperatures between $5600$ and $5900 \,{\rm K}$.

The relations between $\log\,g$ and the proportions of solar-like stars with different chromospheric activity levels are displayed in Figure \ref{fig:ratio}(b).
The proportions of different chromospheric activity levels of solar-like stars appear to be relatively stable in the range of $3.9 < \log\,g< 4.5 \,{\rm dex}$. When $\log\,g>4.5 \,{\rm dex}$, the proportions of active solar-like stars exhibits an increasing trend, whereas the spectral ratios of very inactive, inactive and very active solar-like stars decrease.

\citet{2006AJ....132..161G} and \citet{2017ApJ...848...34H} detected that the distribution of $\log\,R'_{\rm HK}$ varies among stars with different levels of metallicity, and the bimodal distribution \citep{1980PASP...92..385V} is observed in dwarf stars with [Fe/H] greater than $-0.2\,{\rm dex}$.
In the research of \citet{2008A&A...485..571J}, the majority of stars with ${\rm [Fe/H]}>0.1\,{\rm dex}$ were found to be inactive.
In Figure \ref{fig:Rp_HK_PHOENIX_hist}, the bimodal distribution of $\log\,R'_{\rm HK}$ does not exist in our solar-like star sample of LAMOST LRS.
However, as shown in Figure \ref{fig:ratio}(c), when ${\rm [Fe/H]}>0.1\,{\rm dex}$, there is a decrease in the proportions of active solar-like stars.
This decreasing trend ceases and the proportions of active solar-like stars becomes relatively stable when ${\rm [Fe/H]} = 0.3 \,{\rm dex}$.

\section{Summary and Conclusion} \label{sec:conclusion}

In this work, we identify 1,122,495 high-quality LRS spectra of solar-like stars from LAMOST DR8 and provide a database of stellar chromospheric activity parameters based on this spectral sample.
The database contains the stellar chromospheric activity parameters  $S_{\rm tri}$, $S_{\rm MWO}$, $R_{\rm HK}$ and $R'_{\rm HK}$ , as well as their uncertainties. $R_{\rm HK}$ and $R'_{\rm HK}$ are derived from the method in the classic literature (denoted with {\tt\string classic}) and  the method based on the PHOENIX model (denoted with {\tt\string PHOENIX}).
When converting the $S_{\rm MWO}$ to the bolometric calibrated index $R_{\rm HK}$, the $R_{\rm HK,classic}$ values are estimated based on the bolometric factor $C_{\rm cf}$ from \citet{1984A&A...130..353R} and the $K$ factor from \citet{1982A&A...107...31M}, while the $R_{\rm HK,PHOENIX}$ values are derived from the stellar surface flux $\mathcal{F}_{\rm RV}$.
The values of $R_{\rm HK,PHOENIX}$ are approximately $\beta=1.6$ times larger than the values of $R_{\rm HK,classic}$.
For the corresponding photospheric contribution $R_{\rm phot}$, the $R_{\rm phot,classic}$ are deduced based on \citet{1984ApJ...279..763N}, and the $R_{\rm phot,PHOENIX}$ are scaled by $\beta$ times from the $R_{\rm phot,classic}$.
The bolometric and photospheric calibrated chromospheric activity index $R'_{\rm HK}$ is consequently derived by eliminating the photospheric contribution from $R_{\rm HK}$.
Our calculations show that $\log\,R_{\rm HK,PHOENIX}$ and $\log\,R'_{\rm HK,PHOENIX}$ are approximately linearly correlated with $\log\,R_{\rm HK,classic}$ and $\log\,R'_{\rm HK,classic}$, respectively.

We explore the overall properties of stellar chromospheric activity based on the 861,505 solar-like stars in the database. The results show that the median values of $\log\,R'_{\rm HK,PHOENIX}$ with $T_{\rm eff}$ have a minimum at about $T_{\rm eff}=5500\,{\rm K}$, while the dependence of the median values of $\log\,R'_{\rm HK,PHOENIX}$ on $\log\,g$ and [Fe/H] is relatively weak.
The value of solar chromospheric activity index is located at the midpoint of the solar-like star sample. This result from our extensive archive support the view that the dynamo mechanism of solar-like stars is generally consistent with the Sun.
The absence of VP gap in the distribution of chromospheric activity for our solar-like stars could be attributed to three possible factors: 1) a gradual diminishing of chromospheric activity during the evolution of solar-like stars; 2) the influence of different stellar properties on the bimodal distribution of the chromospheric activity within our samples, which should be explored in more detail in the future, or 3) the loss of some information in the spectral profile due to the limited resolution of LAMOST LRS spectra.

We explore the proportions of solar-like stars with different chromospheric activity levels (very active, active, inactive and very inactive).
Based on the values of $\log\,R'_{\rm HK,PHOENIX}$, we can obtain the proportions of very active, active, inactive and very inactive solar-like stars as 1.03\%, 21.68\%, 65.27\% and 12.03\%, respectively. While for the values of $\log\,R'_{\rm HK,classic}$, the proportions are 1.07\%, 24.53\%, 62.98\% and 11.41\%, respectively.
It is observed that the higher the stellar chromospheric activity levels, the narrower the distribution areas in the $T_{\rm eff}$ vs. $\log\,g$, $T_{\rm eff}$ vs. [Fe/H], and [Fe/H] vs. $\log\,g$ parameters spaces.

We further investigate the relation between the proportions of solar-like stars with different chromospheric activity levels (classified by $R'_{\rm HK,PHOENIX}$) and the stellar atmospheric parameters ($T_{\rm eff}$, $\log\,g$ and [Fe/H]).
Based on the proportions of active and very active solar-like stars, it is concluded that the occurrence rate of high levels of chromospheric activity is lower among the stars with effective temperatures between $5600$ and $5900 \,{\rm K}$.
It is found that when $\log\,g>4.5 \,{\rm dex}$, the proportions of active solar-like stars exhibits an increasing trend, whereas the proportions of very inactive, inactive and very active solar-like stars decrease.
It is discovered that there is a decrease in the proportions of active solar-like stars when ${\rm [Fe/H]}>0.1\,{\rm dex}$.
This decreasing trend ceases and the proportions of active solar-like stars becomes relatively stable when ${\rm [Fe/H]} = 0.3 \,{\rm dex}$.

The chromospheric activity database of the LAMOST LRS spectra of solar-like stars provided in this work includes the most commonly used chromospheric activity parameters such as $S_{\rm MWO}$, $R_{\rm HK}$ and $R'_{\rm HK}$.
The relationship between chromospheric activity and other stellar magnetic manifestation (such as stellar rotation period and age) can be further investigated.
Additionally, the database can be used to investigate the relationship between stellar and solar activity for a better understanding of the stellar-solar connection.
The database may also contribute to the discovery of new solar-type stars accommodating potentially habitable exoplanetary systems.

\begin{acknowledgements}

This work is supported by the National Key R\&D Program of China (2019YFA0405000) and the National Natural Science Foundation of China (12073001 and 11973059). W.Z. and J.Z. thank the support of the Anhui Project (Z010118169). H.H. acknowledges the CAS Strategic Pioneer Program on Space Science (XDA15052200) and the B-type Strategic Priority Program of the Chinese Academy of Sciences (XDB41000000). Guoshoujing Telescope (the Large Sky Area Multi-Object Fiber Spectroscopic Telescope, LAMOST) is a National Major Scientific Project built by the Chinese Academy of Sciences. Funding for the project has been provided by the National Development and Reform Commission. LAMOST is operated and managed by the National Astronomical Observatories, Chinese Academy of Sciences.
\end{acknowledgements}

\bibliographystyle{aa}
\bibliography{ref}

\appendix{}
\section{Accuracy of Stellar Parameters}\label{sec:acccuracy_SP}

\begin{figure} 
	\resizebox{\hsize}{!}{\includegraphics{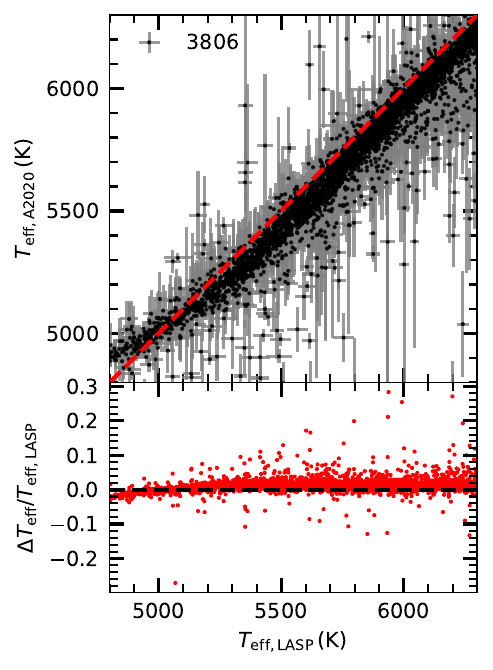}}
	\caption{The distribution of the $T_{\rm eff,LASP}$ values and the $T_{\rm eff,A2020}$ values for 3806 common stars, where $T_{\rm eff,A2020}$ denotes the effective temperature in \citet{2020MNRAS.499.3481A}. The red dashed line represents the case where $T_{\rm eff,LASP}$ equals $T_{\rm eff,A2020}$. Error bars are indicated for data points that have specified uncertainty values. The lower panel displays the distribution of $\Delta T_{\rm eff}/T_{\rm eff,LASP}$ against $T_{\rm eff,LASP}$, where $\Delta T_{\rm eff}=T_{\rm eff,LASP}-T_{\rm eff,A2020}$.
	}
	\label{fig:Teff_compare}
\end{figure}

\begin{figure} 
	\resizebox{\hsize}{!}{\includegraphics{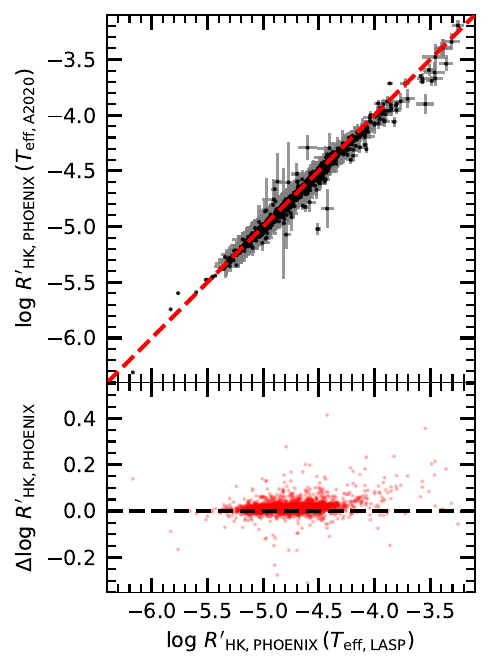}}
	\caption{Scatter plot of $\log\, R'_{\rm HK,PHOENIX}\,(T_{\rm eff,LASP})$ versus $\log\, R'_{\rm HK,PHOENIX}\,(T_{\rm eff,A2020})$ for stars in Figure \ref{fig:Teff_compare}. Error bars are displayed for data points with known uncertainty values. The red dashed line represents the ratio of $\frac{\log\, R'_{\rm HK,PHOENIX}\,(T_{\rm eff,LASP})}{\log\, R'_{\rm HK,PHOENIX}\,(T_{\rm eff,A2020})}=1$. The lower panel shows the distribution of $\Delta \log\, R'_{\rm HK,PHOENIX}$ against $\log\, R'_{\rm HK,PHOENIX}\,(T_{\rm eff,LASP})$.
	}
	\label{fig:Rp_HK_compare}
\end{figure}

We identify 3806 common stars in \citet{2020MNRAS.499.3481A} and compare their effective temperature values with those in our database, as shown in Figure \ref{fig:Teff_compare}.
For $T_{\rm eff}$  in the range of 4800 to 6300 K, the $T_{\rm eff}$ values provided by LASP are approximately consistent with the results in \citet{2020MNRAS.499.3481A}, generally with $\Delta T_{\rm eff}$ less than 120 K.
The $T_{\rm eff,A2020}$ values are obtained from various observation instruments and are taken from the survey with the highest spectral resolution when the same sources were observed in multiple surveys \citep{2020MNRAS.499.3481A}.
Differences in observation instruments and estimation methods would contribute to the discrepancies between $T_{\rm eff,LASP}$ and $T_{\rm eff,A2020}$.
Although there are some differences between $T_{\rm eff,LASP}$ and $T_{\rm eff,A2020}$, the corresponding $\log\, R'_{\rm HK,PHOENIX}$ values estimated based on $T_{\rm eff,LASP}$ and $T_{\rm eff,A2020}$ exhibit approximate consistency as shown in Figure \ref{fig:Rp_HK_compare}, where the $\Delta \log\, R'_{\rm HK,PHOENIX}$ values are generally less than 0.05.

\section{Calibration of Chromospheric Activity Index}\label{sec:calibration}

\begin{figure} 
	\resizebox{\hsize}{!}{\includegraphics{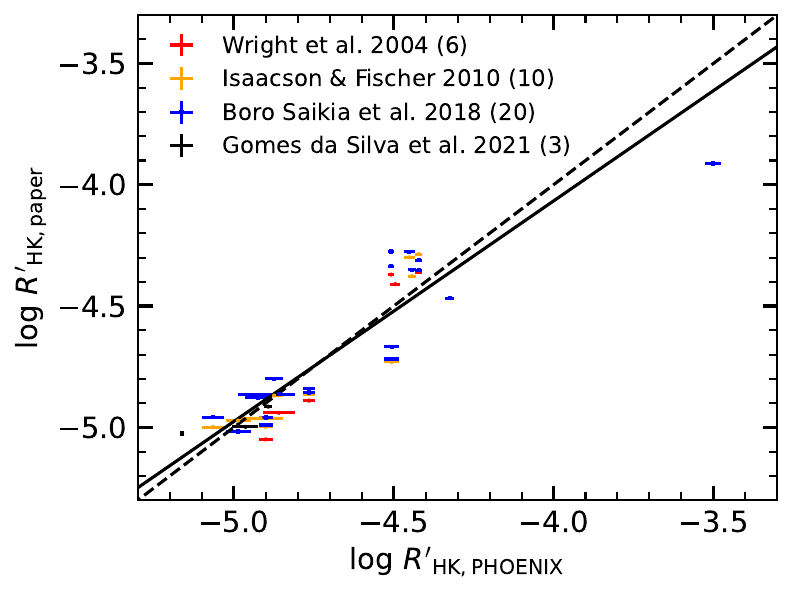}}
	\caption{The distribution of $\log\,R'_{\rm HK,PHOENIX}$ versus $\log\,R'_{\rm HK,paper}$. The dashed line represents the case where the $\log\,R'_{\rm HK,PHOENIX}$ is equal to $\log\,R'_{\rm HK,paper}$, and the solid line is the linear fit between $\log\,R'_{\rm HK,PHOENIX}$ and $\log\,R'_{\rm HK,paper}$. Data points with quantified uncertainty are represented with error bars.
	}
	\label{fig:cross}
\end{figure}

We cross-match the Gaia DR3 source identifier in this paper with stars in \citet{2004ApJS..152..261W}, \citet{2010ApJ...725..875I}, \citet{2018A&A...616A.108B} and ~\citet{2021A&A...646A..77G}, and find out 23 common stars (16 stars in \citealt{2004ApJS..152..261W} and \citealt{2010ApJ...725..875I} are also studied in \citealt{2018A&A...616A.108B}). 
Figure \ref{fig:cross} shows the distribution between $\log\,R'_{\rm HK,PHOENIX}$ and $\log\,R'_{\rm HK,paper}$.
As can be seen in Figure \ref{fig:cross}, our results show an approximate agreement with values from other instruments.
The values used in Figure \ref{fig:cross} are recorded in Table \ref{tab:cross}.

\begin{table*} 
	\centering
	\caption{Common stars for comparing $\log\,R'_{\rm HK,PHOENIX}$ with $\log\,R'_{\rm HK,paper}$ in Figure \ref{fig:cross}. \label{tab:cross}}
	\begin{tabular}{ccccccc}
		\hline \hline
		Gaia DR3 Source ID&$\log\,R'_{\rm HK,PHOENIX}$&$\delta \log\,R'_{\rm HK,PHOENIX}$&$\log\,R'_{\rm HK,paper}$& $\delta \log\,R'_{\rm HK,paper}$&Star Name &Source\\
		\hline
107774198474602368&-4.508228&0.010826&-4.37& &HIP 10679&(1) \\
107774202769886848&-4.494861&0.015467&-4.41& &HIP 10680&(1) \\
2482274463233174016&-4.765261&0.017813&-4.89& &HIP 6712&(1) \\
2559181861327324928&-4.899986&0.021481&-5.05& &HIP 6653&(1) \\
2568947762259903616&-4.859325&0.050142&-4.94& &HIP 9035&(1) \\
2700327032273611264&-4.421454&0.010720&-4.36& &HIP 107107&(1) \\
66838452861270272&-4.442698&0.011885&-4.377& &HD 282954&(2) \\
77161217776670208&-4.505995&0.023880&-4.730& &HD 13357&(2) \\
589826694824322176&-4.986656&0.039343&-4.972& &HIP 46627&(2) \\
2482274463233174016&-4.765261&0.017813&-4.863& &HD 8765&(2) \\
2559181861327324928&-4.899986&0.021481&-4.998& &HD 8648&(2) \\
2664960072535183104&-4.873992&0.027421&-4.869& &HD 219770&(2) \\
2700327032273611264&-4.421454&0.010720&-4.286& &HD 206387&(2) \\
2838213864935858816&-4.451270&0.017570&-4.299& &HD 219498&(2) \\
2868784136478476032&-5.065425&0.033502&-4.999& &HIP 117386&(2) \\
3255358990147896704&-4.916303&0.071081&-4.962& &HD 26257&(2) \\
66838452861270272&-4.442698&0.011885&-4.351& &HD282954&(3) \\
77161217776670208&-4.505995&0.023880&-4.718& &HD13357&(3) \\
77161217776670208&-4.505995&0.023880&-4.667& &HIP10175&(3) \\
107774198474602368&-4.508228&0.010826&-4.275& &HD14082B&(3) \\
107774198474602368&-4.508228&0.010826&-4.336& &HIP10679&(3) \\
164088748804295168&-3.502017&0.025238&-3.912& &HD281691&(3) \\
589826694824322176&-4.986656&0.039343&-5.017& &HIP46627&(3) \\
2482274463233174016&-4.765261&0.017813&-4.841& &HD8765&(3) \\
2482274463233174016&-4.765261&0.017813&-4.857& &HIP6712&(3) \\
2559181861327324928&-4.899986&0.021481&-4.959& &HD8648&(3) \\
2559181861327324928&-4.899986&0.021481&-4.990& &HIP6653&(3) \\
2568947762259903616&-4.859325&0.050142&-4.863& &HIP9035&(3) \\
2664960072535183104&-4.873992&0.027421&-4.800& &HD219770&(3) \\
2700327032273611264&-4.421454&0.010720&-4.311& &HD206387&(3) \\
2700327032273611264&-4.421454&0.010720&-4.353& &HIP107107&(3) \\
2838213864935858816&-4.451270&0.017570&-4.277& &HD219498&(3) \\
2868784136478476032&-5.065425&0.033502&-4.959& &HIP117386&(3) \\
3231423481005237760&-4.924676&0.040069&-4.878& &BD+000873&(3) \\
3255358990147896704&-4.916303&0.071081&-4.865& &HD26257&(3) \\
3314440285393131008&-4.324742&0.014539&-4.468& &HD27990&(3) \\
1153682508388170112&-5.162074&-9999&-5.0264&0.0094&WASP-24&(4) \\
2651240950559225728&-4.892375&0.012469&-4.9144&0.0047&HD 218249&(4) \\
2739638764856168192&-4.963751&0.040451&-4.9979&0.0059&HD 223854&(4) \\
		\hline
	\end{tabular}
	\tablebib{
		(1)~\citet{2004ApJS..152..261W};
		(2)~\citet{2010ApJ...725..875I};
		(3)~\citet{2018A&A...616A.108B};
		(4)~\citet{2021A&A...646A..77G};
	}
\end{table*}

\section{Distribution of Chromospheric Activity Index with multi-observation }\label{sec:CA_multi}

\begin{figure} 
	\resizebox{\hsize}{!}{\includegraphics{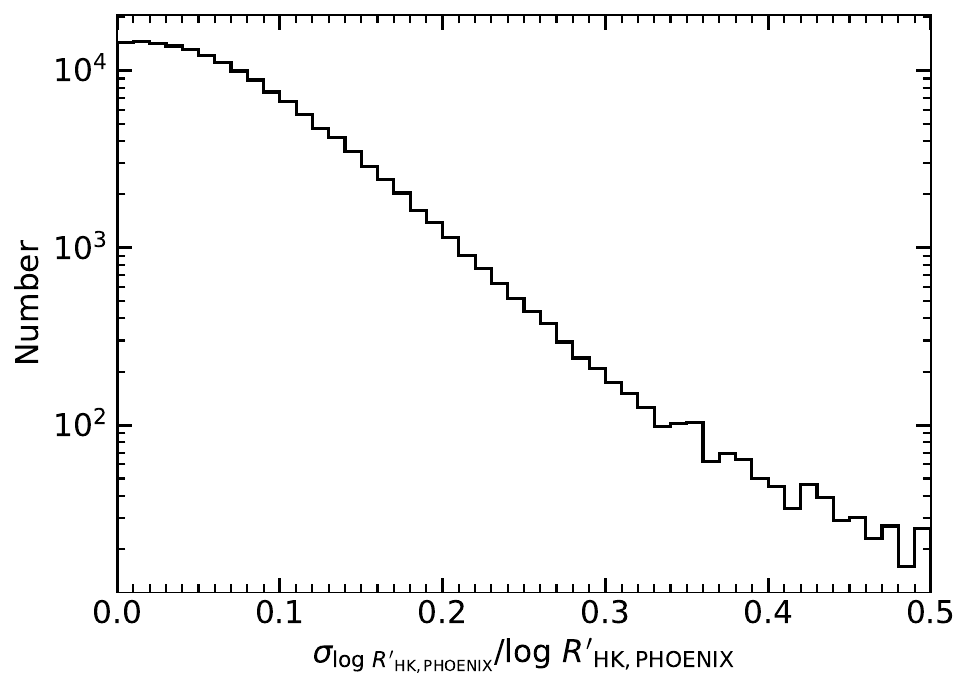}}
	\caption{Histogram of  $\sigma_{\log\,R'_{\rm HK,PHOENIX}}/\log\,R'_{\rm HK,PHOENIX}$ for solar-like stars with more than one observation, where the $\sigma_{\log\,R'_{\rm HK,PHOENIX}}$ represents the standard deviation of $\log\,R'_{\rm HK,PHOENIX}$.
}
	\label{fig:Rp_HK_PHOENIX_std_hist}
\end{figure}

\begin{figure*} 
	\resizebox{\hsize}{!}{\includegraphics{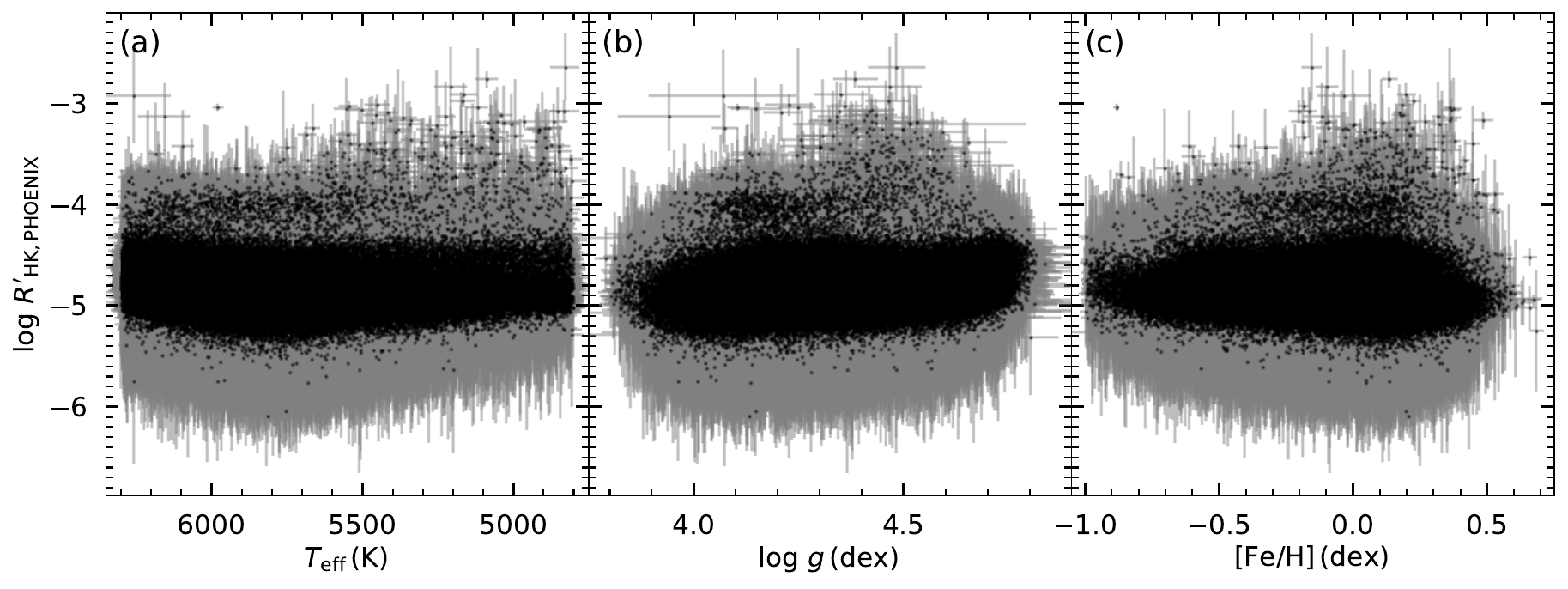}}
	\caption{Distributions of $\log\,R'_{\rm HK,PHOENIX}$ with (a) $T_{\rm eff}$, (b) $\log\,g$ and (c) [Fe/H] for solar-like stars with more than one observation and $\sigma_{\log\,R'_{\rm HK,PHOENIX}}/\log\,R'_{\rm HK,PHOENIX}<0.2$. Error bars represent the standard deviation of $\log\,R'_{\rm HK,PHOENIX}$.
	}
	\label{fig:Rp_HK_PHOENIX_std-teff-logg-feh}
\end{figure*}

The histogram of $\sigma_{\log\,R'_{\rm HK,PHOENIX}}/\log\,R'_{\rm HK,PHOENIX}$ for stars with more than one observation is shown in Figure \ref{fig:Rp_HK_PHOENIX_std_hist}, where the $\sigma_{\log\,R'_{\rm HK,PHOENIX}}$ represents the standard deviation of $\log\,R'_{\rm HK,PHOENIX}$.
The stars with $\sigma_{\log\,R'_{\rm HK,PHOENIX}}/\log\,R'_{\rm HK,PHOENIX}<0.2$ account for 95.5\% of the stars with more than one observation.
Additionally, Figure \ref{fig:Rp_HK_PHOENIX_std-teff-logg-feh} displays the distributions of $\log\,R'_{\rm HK,PHOENIX}$ with $T_{\rm eff}$ (a), $\log\,g$ (b) and [Fe/H] (c) for solar-like stars with more than one observation and $\sigma_{\log\,R'_{\rm HK,PHOENIX}}/\log\,R'_{\rm HK,PHOENIX}<0.2$.
The envelopes in Figure \ref{fig:Rp_HK_PHOENIX_std-teff-logg-feh} are similar to those in Figure \ref{fig:Rp_HK_PHOENIX-teff-logg-feh}.
The distributions of $\log\,R'_{\rm HK,PHOENIX}$ values in (a) $T_{\rm eff}$ vs. $\log\,g$, (b) $T_{\rm eff}$ vs. [Fe/H], and (c) [Fe/H] vs. $\log\,g$ parameter spaces for solar-like stars with more than one observation and $\sigma_{\log\,R'_{\rm HK,PHOENIX}}/\log\,R'_{\rm HK,PHOENIX}<0.2$ are shown in Figure \ref{fig:Rp_HK_PHOENIX-teff-logg-feh_color}.

\begin{figure*} 
	\resizebox{\hsize}{!}{\includegraphics{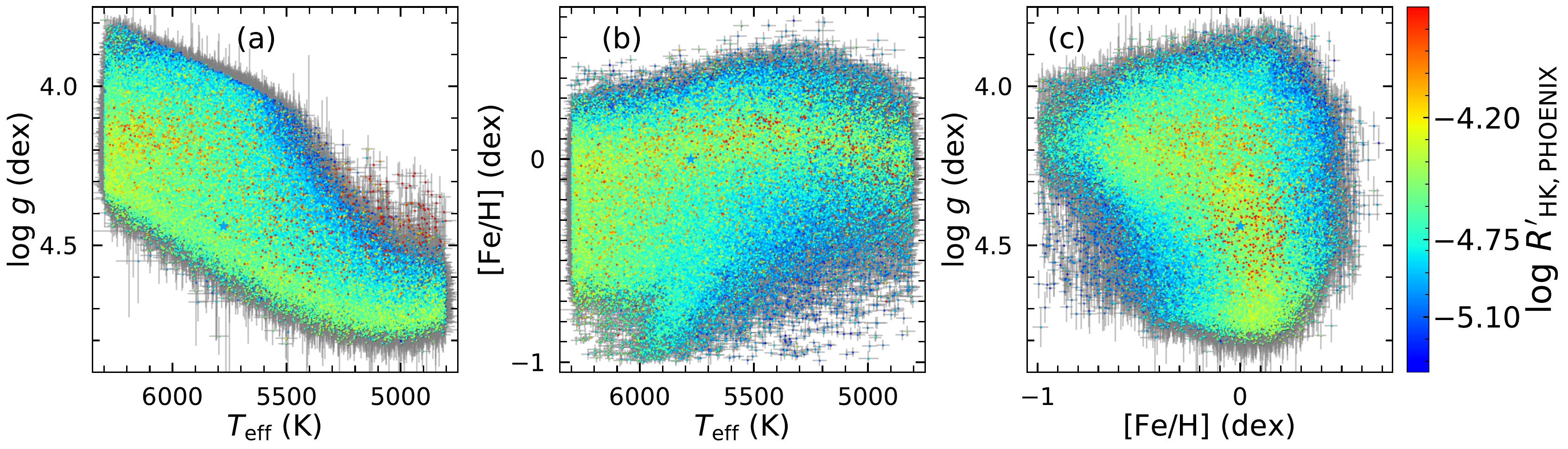}}
	\caption{The distributions of $\log\,R'_{\rm HK,PHOENIX}$ values in (a) $T_{\rm eff}$ vs. $\log\,g$, (b) $T_{\rm eff}$ vs. [Fe/H], and (c) [Fe/H] vs. $\log\,g$ parameter spaces for solar-like stars with more than one observation and $\sigma_{\log\,R'_{\rm HK,PHOENIX}}/\log\,R'_{\rm HK,PHOENIX}<0.2$. The form of this image is similar to Figure \ref{fig:Rp_HK_PHOENIX-teff-logg-feh_color}, where the error bars belong to the bottom of the data point.
	}
	\label{fig:Rp_HK_PHOENIX_std-teff-logg-feh-color}
\end{figure*}

\end{CJK*}
\end{document}